\documentclass[usenatbib]{aa}
\usepackage{graphicx}
\usepackage{multirow}
\usepackage{textcomp}
\usepackage[varg]{txfonts}
\usepackage{xcolor}
\definecolor{cobalt}{rgb}{0.06, 0.2, 0.65}
\usepackage{natbib}
\bibpunct{(}{)}{;}{a}{}{,}
\usepackage{hyperref}    
\hypersetup{
  colorlinks=true,
  citecolor=cobalt,
  linkcolor=cobalt,
  urlcolor=cobalt,
}
\usepackage{float}
\usepackage{cleveref} 
\Crefname{equation}{Eq.}{Eqs.}
\Crefname{figure}{Fig.}{Figs.}
\usepackage{orcidlink}
\usepackage{natbib}
\usepackage{xspace}
\usepackage{exscale, relsize}
\usepackage{siunitx}
\usepackage{caption}

\def\dd{\mathrm{d}}
\def\l{\left}
\def\r{\right}
\def\f{\frac}
\newcommand{\msun}{\ensuremath{\mathrm{M}_\odot}}
\def\CHIMERA{\texttt{\textsc{chimera}}\xspace}
\def\CHIMERAfirst{\texttt{\textsc{chimera \oldstylenums{1.0}}}\xspace}
\def\CHIMERAsecond{\texttt{\textsc{chimera \oldstylenums{2.0}}}\xspace}
\DeclareSIUnit \parsec {pc}
\DeclareSIUnit \years{yrs}

\begin{document}

\title{Accelerating the standard siren method: Improved constraints on modified gravitational-wave propagation with future data}

\author{Matteo Tagliazucchi\inst{1,2,3,}\thanks{Corresponding author: \href{emailto:matteo.tagliazucchi2@unibo.it}{matteo.tagliazucchi2@unibo.it}}\orcidlink{0009-0003-8886-3184}, 
Michele Moresco\inst{1,2}\orcidlink{0000-0002-7616-7136}, Nicola Borghi\inst{1,2,3}\orcidlink{0000-0002-2889-8997}, \and Manfred Fiebig\inst{1}\orcidlink{0009-0003-2729-8708}}

\institute{Dipartimento di Fisica e Astronomia ``Augusto Righi''--Universit\`{a} di Bologna, via Piero Gobetti 93/2, I-40129 Bologna, Italy
\and
INAF - Osservatorio di Astrofisica e Scienza dello Spazio di Bologna, via Piero Gobetti 93/3, I-40129 Bologna, Italy
\and
INFN - Sezione di Bologna, Viale Berti Pichat 6/2, I-40127 Bologna, Italy
}

\abstract{
Gravitational waves (GWs) from compact binary mergers have emerged as one of the most promising probes of cosmology and general relativity (GR).
However, a major challenge in fully exploiting GWs as ``standard sirens'' with current and future GW observatories is developing efficient and robust codes capable of analyzing the increasing data volumes that are, and will be, acquired.
Here, we present \texttt{\textsc{chimera \oldstylenums{2.0}}}, an advanced computational framework for hierarchical Bayesian inference of cosmological, modified gravity, and population hyperparameters using standard sirens and galaxy catalogs.
This upgrade introduces novel GPU-accelerated algorithms to estimate the hierarchical likelihood, enabling the analysis of thousands of events - crucial for next-generation experiments - and includes the two-parameter ($\Xi_0-n$) modified GW propagation model, where $\Xi_0$ governs the amplitude of the modification ($\Xi_0 = 1$ corresponds to GR).
Using \CHIMERAsecond, we forecast cosmological and modified GW propagation constraints for a scenario similar to the future LIGO-Virgo-KAGRA O5 run.
We analyze three binary black-hole populations of 300 events at S/N>20, each with a different value of $\Xi_0$: 0.6, 1 (corresponding to GR), and 1.8. 
Multiple analyses were performed each catalog, comprising a population of approximately 5000 events, thanks to \CHIMERAsecond, which is 10-1000 times faster depending on the settings and catalog size.
We jointly infer cosmological, modified GW propagation, and population hyperparameters.
With spectroscopic galaxy catalogs, the fiducial $\Xi_0$ is recovered with a precision of $22\%$, $7.5\%$, and $10\%$ for $\Xi_0=0.6$, $1$, and $1.8$, respectively; while the precision on $H_0$ is $2-7$ times worse than when $\Xi_0$ is not inferred.
Finally, in the case of photometric redshifts the constraints degrade on average by 3.5 times in all cases, underscoring the importance of future spectroscopic surveys in maximizing the constraining power of standard sirens.
}

\keywords{gravitation - gravitational waves - methods: data analysis – methods: statistical – cosmological parameters – cosmology: observations}

\titlerunning{Accelerating the standard siren method: improved constraint on modified GW propagation}
\authorrunning{M. Tagliazucchi et al.}

\date{Received 28 March 2025 / Accepted 26 August 2025}

\maketitle

\section{Introduction}

The first direct detection of gravitational waves (GWs) with GW150914 \citep{LIGOScientific:2016aoc} marked the beginning of a new era in astronomy and cosmology.
GWs from compact binary coalescences (CBC) can be used as ``standard sirens'' as they provide a direct measurement of the luminosity distance to the source, without the need for any intermediate calibrator \citep{Schutz:1986gp, Holz:2005df, Moresco:2022phi}.
By combining standard sirens with information on the redshift of the source, it becomes possible to study the expansion history of the Universe through the electromagnetic luminosity distance-redshift relation:
\begin{equation}\label{eq:dL-em}
    d^{\rm em}_L(z; \boldsymbol{\lambda}_c) = (1+z)\int_0^z \f{c\dd z}{H(z;\boldsymbol{\lambda}_c)} ,
\end{equation}
where we assumed a flat geometry of the Universe, and $\boldsymbol{\lambda}_c$ represents cosmological parameters such as the Hubble constant, $H_0$, and the matter energy density, $\Omega_{m}$.
Standard sirens are, and will increasingly become, a powerful tool for addressing tensions in cosmological parameters, such as the Hubble tension, that have emerged in the recent era of precision cosmology \citep{Verde:2019ivm, Moresco:2022phi}.
Interestingly, standard sirens also allow us to constrain possible deviations in the propagation of gravitational and electromagnetic signals, providing an additional test of general relativity (GR).
The multimessenger observation of the binary neutron star coalescence GW170817 \citep{LIGOScientific:2017zic, LIGOScientific:2017vwq} placed a stringent limit on the difference between the propagation speed of gravitational and electromagnetic waves, $|c_{\rm gw} - c|/c < \mathcal{O}\l(10^{-15}\r)$, ruling out many modified-gravity (MG) models.
All MG models consistent with this constraint introduce a friction term in the propagation of tensor modes at cosmological scales, which is absent in GR \citep{LISACosmologyWorkingGroup:2019mwx}. 
This friction term modifies the effective distance to which standard sirens are sensitive, making them an effective tool to probe MG models \citep{Lombriser:2015sxa, Saltas:2014dha, Nishizawa:2017nef, Arai:2017hxj, Amendola:2017ovw, Belgacem:2017ihm, Belgacem:2018lbp}. 
A common way to parameterize the friction term involves two parameters $\boldsymbol{\lambda}_{mg} = (\Xi_0,n)$.
In this parameterization, the effective luminosity distance measured by standard sirens is given by \citep{Belgacem:2017ihm}
\begin{equation}
\begin{aligned}
    d_L^{\rm gw}(z; \boldsymbol{\lambda}_c, \boldsymbol{\lambda}_{mg}) = d_L^{\rm em}(z; \boldsymbol{\lambda}_c)\; \Xi(z;\boldsymbol{\lambda}_{mg}) = \\ 
    d_L^{\rm em}(z; \boldsymbol{\lambda}_c)\; \left[\Xi_0 + \frac{1 - \Xi_0}{(1 + z)^n}\right]
    \label{eq:dL-mg}
\end{aligned}
\end{equation}
where $\Xi(z;\boldsymbol{\lambda}_{mg})$ quantifies the deviation from the standard electromagnetic distance \Cref{eq:dL-em} at each redshift.
This parameterization can represent several scalar-tensor theories of the Horndeski class, including the Brans-Dicke, $f(R)$, covariant-Galileon, and minimal-acceleration models (see Table 1 in \cite{Belgacem:2018lbp} for a summary). It can also be connected with nonlocal gravity theories \citep{Belgacem:2019lwx, Belgacem:2020pdz}.

One of the main challenges to exploiting the standard siren method for cosmological purposes is the perfect degeneracy between the binary redshift and the chirp mass.
This degeneracy, due to the scale invariance of GR \citep{2022hgwa.bookE..48M, Mastrogiovanni:2024mqc}, can be broken if a physical effect imprints a known scale on the GW signal or if external redshift information is provided.
An example of a standard siren technique that determines the redshift by leveraging physical scales in the GW signal is the ``spectral-siren'' method, which exploits features in the source-frame mass distribution of CBC populations to statistically infer the redshift \citep{Chernoff:1993th, Taylor:2011fs, Farr:2019twy, Mastrogiovanni:2021wsd, Mukherjee:2021rtw, Mancarella:2021ecn, Karathanasis:2022rtr, Chen:2024gdn}.
Examples of such features include those observed in the binary black hole (BBH) mass distribution inferred from LIGO-Virgo-KAGRA (LVK) data up to the observing run O3, which reveals an overdensity at approximately $35 \,\msun$ and a steep decrease after $80 \,\msun$ \citep{KAGRA:2021duu}.
There are two main ways to include external redshift information in the inference process.
The most intuitive is to use, if identified, the redshift of the electromagnetic counterpart when the standard siren is ``bright'' \citep{Holz:2005df, Nissanke:2009kt}; this requires the presence of an electromagnetic phenomenon associated with the CBC, such as a kilonova with the merger of a binary neutron star.
The other method, also known as the ``galaxy catalog method'', requires the GW luminosity-distance probability distribution to be statistically combined with a redshift prior distribution constructed from a catalog of potential hosts, usually a galaxy catalog \citep{Schutz:1986gp, DelPozzo:2011vcw, LIGOScientific:2018gmd, Gray:2019ksv, Gair:2022zsa}.
All three approaches rely on assumptions about the underlying population, whether it is the host galaxy for bright sirens, the redshift prior in the galaxy-catalog method, or the mass distribution in the spectral-siren case. 
Various pipelines have been developed to implement the spectral-siren method, namely \texttt{POPMODELS} \citep{popmodels2017}, \texttt{GWPopulation} \citep{Talbot:2019okv}, \texttt{MGCosmoPop} \citep{Mancarella:2021ecn}, \texttt{\textsc{icarogw} \oldstylenums{1.0}} \citep{Mastrogiovanni:2021wsd}, \texttt{SODAPOP} \citep{sodapop2021}, \texttt{GWInferno} \citep{Edelman:2022ydv}, and in \cite{Chen:2024gdn}.
More recently, a differentiable pipeline that samples the full hierarchical population posterior was released \citep{Mancarella:2025uat}.
For the galaxy-catalog method, tools such as \texttt{DarkSirensStat} \citep{Finke:2021aom}, \texttt{\textsc{gwcosmo}} \citep{Gray:2023wgj}, and - specifically for LISA - \texttt{cosmolisa} \citep{Laghi:2021pqk} are available.
In recent years, new codes have been developed that unify the different methods in a unique Bayesian framework, jointly inferring cosmological and MG hyperparameters with population ones to provide robust constraints: \texttt{\textsc{icarogw} \oldstylenums{2.0}} \citep{Mastrogiovanni:2023emh, Mastrogiovanni:2023zbw}, \texttt{\textsc{gwcosmo}} \citep{Gray:2023wgj}, and \CHIMERA \citep{Borghi:2023opd} (referred to as \CHIMERAfirst from now on).

Constraints on the Hubble constant $H_0$ from standard sirens have been obtained using the publicly available GWTC-3 data \citep{LIGOScientific:2018mvr, LIGOScientific:2020ibl, KAGRA:2021vkt}. 
These include measurements from the bright siren GW170817 \citep{LIGOScientific:2017adf, Palmese:2023beh}, as well as dark sirens in GWTC-3 combined with the GLADE+ \citep{Dalya:2021ewn} galaxy catalog \citep{LIGOScientific:2018gmd, LIGOScientific:2019zcs, Finke:2021aom, LIGOScientific:2021aug, Mastrogiovanni:2023emh, Gray:2023wgj}, the DES survey \citep{DES:2019ccw, DES:2020nay}, the DELVE survey \cite{Alfradique:2023giv}, or DESI \citep{DESI:2023fij}. 
More recent work has extended these analyses to dark sirens from the O4 run \citep{Bom:2024afj}.
Similarly, constraints on $\Xi_0$ have been derived from GWTC-3 data.
For example, \cite{Finke:2021aom} derived $\Xi_0 = 2.1^{+3.2}_{-2.1}$ using dark sirens and GLADE+, while \cite{Mancarella:2021ecn} constrained  $\Xi_0$ with $58\%$ uncertainty using the spectral siren method. 
More recently, joint constraints on $H_0$ and $\Xi_0$ were obtained by combining the spectral siren and galaxy catalog methods, yielding am uncertainty of $73\%$ on $\Xi_0$ and $58\%$ on $H_0$ using 42 BBH events of GWTC-3 and the GLADE+ galaxy catalog \citep{Chen:2023wpj}. 
Preliminary forecasts on future constraints for MG and cosmological parameters have been explored using spectral sirens.
For instance, \cite{Leyde:2022orh} predicts a $20\%-30\%$ measurement of $\Xi_0$ using about 400 detections at the design LVK O5 sensitivity in a GR scenario. Forecasts using bright sirens \citep[e.g.,][]{Niu:2021nic, Chen:2024xkv, Colangeli:2025bnb} or alternative approaches, such as the GW galaxies cross-correlation \citep[e.g.,][]{Mukherjee:2020mha, Afroz:2023ndy,Afroz:2024joi}, have also been investigated.

In general, standard siren codes are computationally limited by the number of GW events they can process.
To forecast how next-generation interferometers, such as the Einstein Telescope \citep{Branchesi:2023mws, Abac:2025saz}, Cosmic Explorer \citep{Reitze:2019iox}, and LISA \citep{Colpi:2024xhw}, will constrain cosmology and modified GW propagation, it is necessary to improve existing pipelines, as these experiments will detect up to $10^5$ BBHs per year.

This paper has two primary aims.
First, we present \CHIMERAsecond, an enhanced version of \CHIMERAfirst that can handle up to tens of thousands of GW events, a critical step toward next-generation detector data.
This is achieved by introducing three different kernel-density-estimate (KDE) algorithms, which form the backbone of the pipeline.
These KDEs provide better flexibility and fully leverage GPU acceleration for optimal performance.
In a future study, we will test and validate this code against other existing pipelines through a blinded-mock-data challenge.
This work will assess how the computational cost of current codes scales with the number of events and identify potential systematic effects in the implemented algorithms.
Second, we used this upgraded code, that includes MG models, to forecast joint cosmological and MG constraints for the first time for the future O5 observing run of the LVK collaboration using the combination of spectral-siren and galaxy-catalog methods.

This paper is organized as follows.
\Cref{sec:methods} outlines the statistical framework of combined standard siren methods and the enhanced numerical implementation of \CHIMERAsecond, comparing it with \CHIMERAfirst in terms of results and computational efficiency.
\Cref{sec:data} describes the mock catalogs used to forecast O5-like constraints on cosmology and modified GW propagation.
We studied three different scenarios: one in which there is no MG, one with a $\Xi_0$ greater than 1, and one with $\Xi_0<1$. 
Finally, \Cref{sec:results} presents the results. 

\section{Methods}\label{sec:methods}

\subsection{Statistical framework} 

Standard sirens as cosmological probes provide constraints on the hyperparameters ($\boldsymbol{\Lambda}$) describing the properties of the CBC population and the underlying cosmological model from a set of observations drawn from that population.
Observations are incomplete due to selection biases in GW interferometers and noisy because the single event parameters in detector-frame $\left\{\boldsymbol{\theta}^\dd_i\right\}_{i=1}^{N_{\rm obs}}$ (e.g., binary redshifted masses, spins, luminosity distance, and localization area) are inferred from a set of observations $\left\{\boldsymbol{d}_i\right\}_{i = 1}^{N_{\rm obs}}$ including measurement uncertainties.
The correct Bayesian framework that accounts for selection effects and noisy observations is a hierarchical inhomogeneous Poisson process, described by the hyper-likelihood \citep{Loredo:2004nn, Mandel:2018mve, Vitale:2020aaz}:
\begin{align}
    &\mathcal{L}\l( \{\boldsymbol{d}_i\}_{i = 1}^{N_{\rm obs}}  \mid \boldsymbol{\Lambda}\r)
     \propto \f{1}{[ \xi (\boldsymbol{\Lambda})]^{N_{\rm obs}}} \prod_{i = 1}^{N_{\rm obs}} \int \dd \boldsymbol{\theta}^\dd_i \, \f{p_{\rm gw}(\boldsymbol{\theta}^\dd_{i} \mid \boldsymbol{d}_i)}{\pi(\boldsymbol{\theta}^{\dd}_i)}  p_{\rm pop}(\boldsymbol{\theta}^\dd_{i} \mid \boldsymbol{\Lambda}), \label{eq:hyperlike}\\
    & \xi (\boldsymbol{\Lambda}) 
    =  \int \dd \boldsymbol{\theta}^\dd P_{\rm det}(\boldsymbol{\theta}^\dd) p_{\rm pop}(\boldsymbol{\theta}^\dd \mid \boldsymbol{\Lambda}) \label{eq:bias}.
\end{align}
$\xi(\boldsymbol{\Lambda})$, representing the fraction of population that can be detected, corrects the Malmquist bias due to selection effects and depends on the probability, $P_{\rm det}(\boldsymbol{\theta}^\dd)$, of detecting a source with detector-frame parameters $\boldsymbol{\theta}^\dd$:
\begin{equation}\label{eq:p-det}
    P_{\rm det}\l(\boldsymbol{\theta}^\dd \r) = \int_{\boldsymbol{d}\in\text{detectable}} \dd \boldsymbol{d} \,\f{p_{\rm gw}(\boldsymbol{\theta}^\dd \mid \boldsymbol{d})}{\pi(\boldsymbol{\theta}^{\dd})}.
\end{equation}
The term $p_{\rm gw}$ is the posterior distribution for the detector-frame parameters given the data, and it accounts for the measurement uncertainties.
The prior distribution, $\pi$, which appears in the denominator of the integrand, converts this posterior to its corresponding likelihood.
The population prior, $p_{\rm pop}$,gives the probability of drawing an event with parameters $\boldsymbol{\theta}^\dd$ from a population described by hyperparameters, $\boldsymbol{\Lambda}$. 

The term $p_{\rm pop}$ is usually modeled in terms of source frame parameters $\boldsymbol{\theta}$.
Here, we neglected spins and focused only on binary masses, $m_{1,2}$, redshift, $z$, and sky position, $\hat{\Omega}$.
The cosmological and MG hyperparameter $(\boldsymbol{\lambda}_c, \boldsymbol{\lambda}_{mg})$ map between detector- and source-frame parameters. 
Specifically, they convert the measured luminosity distance into the corresponding redshift via \Cref{eq:dL-mg}.
The redshift is then used to transform the binary masses in the source frame as $m^\dd_{1,2} = (1+z)\,m_{1,2}$.
By assuming that the mass distribution does not evolve with redshift, the source-frame population prior can be factorized as
\begin{equation}\label{eq:pop-general}
    p_{\rm pop}(\boldsymbol{\theta} \mid \boldsymbol{\Lambda}) = p(m_1, m_2 \mid \boldsymbol{\lambda}_m) \, p_{\rm cbc}(z, \hat\Omega \mid \boldsymbol{\lambda}_c,  \boldsymbol{\lambda}_r),
\end{equation}
where the additional hyperparameters $\boldsymbol{\lambda}_m$ and $\boldsymbol{\lambda}_r$ describe the mass and merger-rate evolution distributions, respectively. 
The set of all different hyperparameters is denoted by $\boldsymbol{\Lambda} = \{\boldsymbol{\lambda}_c, \boldsymbol{\lambda}_{mg}, \boldsymbol{\lambda}_m, \boldsymbol{\lambda}_r\}$.
The first term in \Cref{eq:pop-general} describes the mass distribution of CBCs, while $p_{\rm cbc}$ is the probability of having a CBC at redshift, $z$, and sky position $\hat{\Omega}$.
The latter probability can be written as the probability of having a host galaxy at $(z,\hat{\Omega})$, denoted by $p_{\rm gal}(z,\hat{\Omega} \mid \boldsymbol{\lambda}_c)$, times a function describing the merger-rate redshift evolution, denoted by $\psi(z ; \boldsymbol{\lambda}_r)$.
The distribution $p_{\rm gal}$ is expressed as~\citep{Chen:2017rfc, Finke:2021aom}: 
\begin{equation}
    p_{\rm gal}(z,\hat{\Omega} \mid \boldsymbol{\lambda}_c) = f_R p_{\rm cat}(z,\hat{\Omega}) + (1-f_R) p_{\rm miss}(z,\hat{\Omega}),
\end{equation}
where
\begin{align}
    p_{\rm cat}(z,\hat{\Omega}) \propto \sum_g w_g  \delta(\Omega - \Omega_g)     \f{\mathcal{N}(z; \tilde z_g, \tilde \sigma_{z,g}) \f{\dd V_c}{\dd z}(z; \boldsymbol{\lambda}_c )}{\int \dd z \mathcal{N}(z; \tilde z_g, \tilde \sigma_{z,g}) \f{\dd V_c}{\dd z}(z; \boldsymbol{\lambda}_c )}, \label{eq:p-cat}\\
    p_{\rm miss}(z,\hat{\Omega}) = \f{1-P_{\rm compl}(z,\hat{\Omega})}{1-f_R}\f{1}{V_c(z_{\rm max}; \boldsymbol{\lambda}_c)}\f{\dd V_c}{\dd z }(z;\boldsymbol{\lambda}_c), \label{eq:p-miss}\\
    f_R = \f{1}{V_c(z_{\rm max}; \boldsymbol{\lambda}_c)} \int \dd V_c \, P_{\rm compl}(z,\hat{\Omega}) \label{eq:f-R}.
\end{align}
In this framework, $p_{\rm cat}$ is the probability distribution derived from the galaxy catalog and is modeled as a sum of Gaussian distributions centered on the measured galaxy redshifts ($\tilde z_g$), with standard deviations equal to the measurement uncertainties ($\tilde \sigma_g)$.
Each Gaussian is multiplied by a prior distribution, assumed to be uniform in comoving volume in the absence of additional information \citep{Gair:2022zsa}.
The galaxy positions on the sky are assumed to be known without uncertainty.
The term $p_{\rm miss}$ accounts for the missing galaxies, and depends on $P_{\rm compl}$, the probability of missing a galaxy at $(z,\hat{\Omega})$, as well as on assumptions on how missing galaxies are distributed.
In \Cref{eq:p-miss}, they are assumed to be homogeneously distributed.
However, more accurate prescriptions that account for galaxy clustering properties have recently been proposed \citep{Finke:2021aom, Dalang:2023ehp, Leyde:2024tov, Dalang:2024gfk}.
Finally, $f_R$ \eqref{eq:f-R} is the galaxy-catalog completeness fraction. 

To use the source-frame population prior \eqref{eq:pop-general} in \Cref{eq:hyperlike} it is necessary to take into account the Jacobian of the transformation $\boldsymbol{\theta}^\dd \to \boldsymbol{\theta}(\boldsymbol{\theta}^\dd;\boldsymbol{\lambda}_c, \boldsymbol{\lambda}_{mg})$:
\begin{equation}\label{eq:jacobian}
    p_{\rm pop}(\boldsymbol{\theta}^\dd \mid \boldsymbol{\Lambda}) \propto p_{\rm pop}(\boldsymbol{\theta} \mid \boldsymbol{\Lambda}) \times \l|\f{\dd \boldsymbol{\theta}^\dd}{\dd\boldsymbol{\theta}} \r|^{-1} = \f{p_{\rm pop}(\boldsymbol{\theta} \mid \boldsymbol{\Lambda})}{(1+z)^3 \, \l|\f{\partial d_L}{\partial z}(z; \boldsymbol{\lambda}_c, \boldsymbol{\lambda}_{mg}) \r|},
\end{equation}
where one $(1+z)$ factor expresses the conversion of the merger rate from the source to the detector-frame, and the other two come from the redshifting of binary masses, and the term $\l|\partial d_L/\partial z\r|$ is due to the luminosity distance-redshift conversion.

\subsection{\texorpdfstring{\texttt{\textsc{chimera \oldstylenums{1.0}}}}{chimera 1.0}}
In this section, we describe the numerical implementation of \CHIMERAfirst and its main computational bottlenecks.

The term related to the selection bias $\xi(\boldsymbol{\Lambda})$ \eqref{eq:bias} is estimated using a set of $N_{\rm inj}$ simulated events ({\em \emph{injections}}) with parameters, $\{\boldsymbol{\theta}^\dd_j\}_{j=1}^{N_{\rm inj}}$, that span the detectable parameter space.
The integral in \Cref{eq:bias} is approximated by the following Monte Carlo summation over the set of injections \citep{Talbot:2019okv, Thrane:2018qnx, Essick:2022ojx}:
\begin{equation}
    \xi(\boldsymbol{\Lambda}) \approx \f{1}{N_\mathrm{inj}} \sum_{j=1}^{N_\mathrm{det}}  \f{p_{\rm pop}(\boldsymbol{\theta}_j \mid \boldsymbol{\Lambda})}{(1+z_j)^3 \, \l|\f{\partial d_L}{\partial z}(z_j; \boldsymbol{\lambda_c}, \boldsymbol{\lambda}_{mg}) \r|} \equiv \f{1}{N_\mathrm{inj}} \sum_{j=1}^{N_\mathrm{det}} s_j,
\end{equation}
where we inserted \Cref{eq:p-det,eq:jacobian} into \Cref{eq:bias}.
We checked the numerical stability of the previous finite Monte Carlo summation by requiring that the ``effective'' number of injections, defined as in \cite{Farr:2019rap}
\begin{equation}
    N^{\rm inj}_{\rm eff} = \l[ \sum\limits_{j = 0}^{N_{\rm det}} s_j\r]^2 \times \l[ \sum\limits_{j = 0}^{N_{\rm det}}s_j^2-\f{1}{N_{\rm inj}}\l( \sum\limits_{j = 0}^{N_{\rm det}} s_j\r)^2\r]^{-1},
\end{equation}
is larger than $5 N_{\rm det}$.

The $N_{\rm obs}$ integrals appearing in the numerator of \Cref{eq:hyperlike}, which we denote as $I_i(\boldsymbol{d}_i \mid \boldsymbol{\Lambda})$, are calculated in the source frame over the three-dimensional volume defined by the GW event localization area, $\delta\hat{\Omega}_i$, and redshift interval, $\delta z_i(\boldsymbol{\lambda}_c, \boldsymbol{\lambda}_{mg})$:
\begin{align}\label{eq:like-integrals}
    I_i(\boldsymbol{d}_i\mid\boldsymbol{\Lambda}) = \int_{\delta\Omega_i \times \delta z_i(\boldsymbol{\lambda}_c, \boldsymbol{\lambda}_{mg})} \dd^2\hat{\Omega}\dd z \, \times \nonumber \\ \mathcal{K}_{\mathrm{gw}, i}(z,\hat{\Omega} \mid \boldsymbol{d}_i, \boldsymbol{\lambda}_c, \boldsymbol{\lambda}_{mg}, \boldsymbol{\lambda}_{m})  \f{p_{\rm cbc}(z,\hat{\Omega} \mid \boldsymbol{\lambda}_c, \boldsymbol{\lambda}_r)}{(1+z)^3 \l|\f{\partial d_L}{\partial z}(z;\boldsymbol{\lambda}_c, \boldsymbol{\lambda}_{mg})\r|}.
\end{align}
Here, we used the transformation $\dd \boldsymbol{\theta}^\dd_i \, p_{\rm gw}(\boldsymbol{\theta}^\dd_i \mid \boldsymbol{d}_i) = \dd \boldsymbol{\theta}_i \, p_{\rm gw}(\boldsymbol{\theta}_i \mid \boldsymbol{d}_i; \boldsymbol{\lambda}_c, \boldsymbol{\lambda}_{mg})$.
The integrand in \Cref{eq:like-integrals} includes terms from the population prior \eqref{eq:pop-general} and the Jacobian \eqref{eq:jacobian} that depend only on the redshift and/or sky position.
These terms are multiplied by the GW event kernel, $\mathcal{K}_{\mathrm{gw}, i}$, defined as the GW posterior weighted by the mass distribution and detector-frame parameter priors, marginalized over $m_{1,2}$ and evaluated on the integration volume $(z,\hat{\Omega})$: 
\begin{align}\label{eq:gw-kernel}
    \mathcal{K}_{\mathrm{gw}, i}(z,\hat{\Omega} \mid \boldsymbol{d}_i, \boldsymbol{\lambda}_c, \boldsymbol{\lambda}_{mg}, \boldsymbol{\lambda}_{m}) =  \int \dd m_{1,i} \dd m_{2,i}\, \times \nonumber \\ \l. p_{\rm gw}(m_{1,i}, m_{2,i}, z_i, \hat{\Omega}_i \mid \boldsymbol{d}_i;\boldsymbol{\lambda}_c, \boldsymbol{\lambda}_{mg}) \f{p(m_{1,i}, m_{2,i} \mid \boldsymbol{\lambda}_m)}{\pi(m^\dd_{1,i} ,m^\dd_{2,i}) \, \pi(d_{L,i})}\r|_{(z,\hat{\Omega})}.
\end{align}

Here, we discuss the "3D kernel." The GW kernel \eqref{eq:gw-kernel} is approximated using a weighted KDE built using $N_{\rm s}$ posterior estimate (PE) samples drawn from $p_{\rm gw}$. 
This algorithm employs a Gaussian kernel and a 3D training dataset:
\begin{align}
     \mathcal{K}_{\mathrm{gw}, i}(z,\hat{\Omega} \mid \boldsymbol{d}_i, \boldsymbol{\lambda}_c, \boldsymbol{\lambda}_{mg}, \boldsymbol{\lambda}_m) \approx \nonumber \\ 
     \l. \text{KDE}\l[\l\{(z^j_i,\hat{\Omega}^j_i) \l| w^j_i = \f{p(m^j_{1,i}, m^j_{2,i} \mid \boldsymbol{\lambda}_m)}{\pi(m^{\dd,j}_{1,i}, m^{\dd,j}_{2,i}) \, \pi(d^j_{L,i})}\r\}_{j=1}^{N_{\rm s}}\r. \r]\r|_{(z,\hat{\Omega})}.
\end{align}
Here, the index $j$ refers to the PE samples of the $i$-th event.
This KDE is evaluated on a 3D grid that approximates the integration domain of \Cref{eq:like-integrals}; it is constructed as follows:
\begin{enumerate}[a.]
    \item Divide the 2D localization area, $\delta\hat{\Omega}_i$, into $N^i_{\rm pix}$ equal-area pixels (see top panel of \Cref{fig:kde-comparison}), as first proposed in \cite{Gray:2021sew}. 
    \item Discretize the redshift interval, $\delta z_i(\boldsymbol{\lambda}_c, \boldsymbol{\lambda}_{mg})$, into $N_{\rm z}$ equally spaced points, ensuring that the grid covers all possible redshifts given the prior on cosmological and MG hyperparameters. This allows us to significantly reduce the computational cost by computing the $p_{\rm cat}$ term only once before the inference.
    \item Repeat the redshift grid $N_{\rm pix}$ times and duplicate the pixel center coordinates $N_{\rm z}$ times to match the redshifts within each pixel.
\end{enumerate}

In \CHIMERAfirst, the KDE evaluation is the main computational bottleneck, accounting for approximately $80\%$ of the total time required for a complete population fit in a scenario involving 100 GW events - with 5000 PE samples per event - and $2\times 10^7$ injections.
The cumulative computational time for the KDE evaluation scales as follows:
\begin{equation}
    t_{\rm KDE} \sim \mathcal{O}\l( N_{\rm obs} \times N_{\rm z} \times N_{\rm pix} \times N_{\rm s}  \r),
\end{equation}
where $N_{\rm z}, N_{\rm pix}$ are the resolutions of redshift and localization area integration volumes, respectively.
In the above scenario, the population fit required $10^4$ CPU hours using the \texttt{emcee} sampler \citep{Foreman-Mackey:2012any} with 50 walkers parallelized  across 25 Intel CPUs (2.0 GHz, 1 core per CPU) on a HPC facility.
Due to the linear dependence on the number of GW events, the KDE bottleneck imposes a significant computational challenge for future next-generation interferometers that will detect hundreds of thousands of GW events.
For example, \CHIMERAfirst analyzes around 1000 events in more than a month of CPU time - a feasible but increasingly impractical burden for even larger catalogs.
This computational time cannot be reduced by simply allocating more CPUs due to inherent limitations in most of the sampling algorithms. Instead, more efficient algorithms and hardware accelerators as the GPU were preferred in order to reduce the computational burden of the likelihood evaluation.

\subsection{Enhanced numerical implementation} 

To overcome the computational limitations of \CHIMERAfirst, we introduced a new release featuring a redesigned pipeline architecture and advanced KDE algorithms.

The new code release, \CHIMERAsecond\footnote{The code is publicly available at \url{https://github.com/CosmoStatGW}}, is fully implemented in the \texttt{JAX} framework \citep{jax2018github}.
\texttt{JAX} is a high-performance Python library for numerical computing and machine learning that combines NumPy-like syntax with just-in-time compilation, automatic differentiation, and GPU or TPU acceleration. 
This enables \CHIMERAsecond to leverage gradient-based Markov chain Monte Carlo (MCMC) algorithms, such as Hamiltonian MCMC algorithms.
Population and cosmological models are constructed using \texttt{equinox} \citep{kidger2021equinox}, which ensures data structures are compatible as \texttt{JAX pytrees} and supports hyperparameter vectorization. Additionally, \CHIMERAsecond now includes extended cosmological models, such as curved and evolving dark-energy universes.

\begin{figure}
    \centering
    \resizebox{\hsize}{!}{\includegraphics{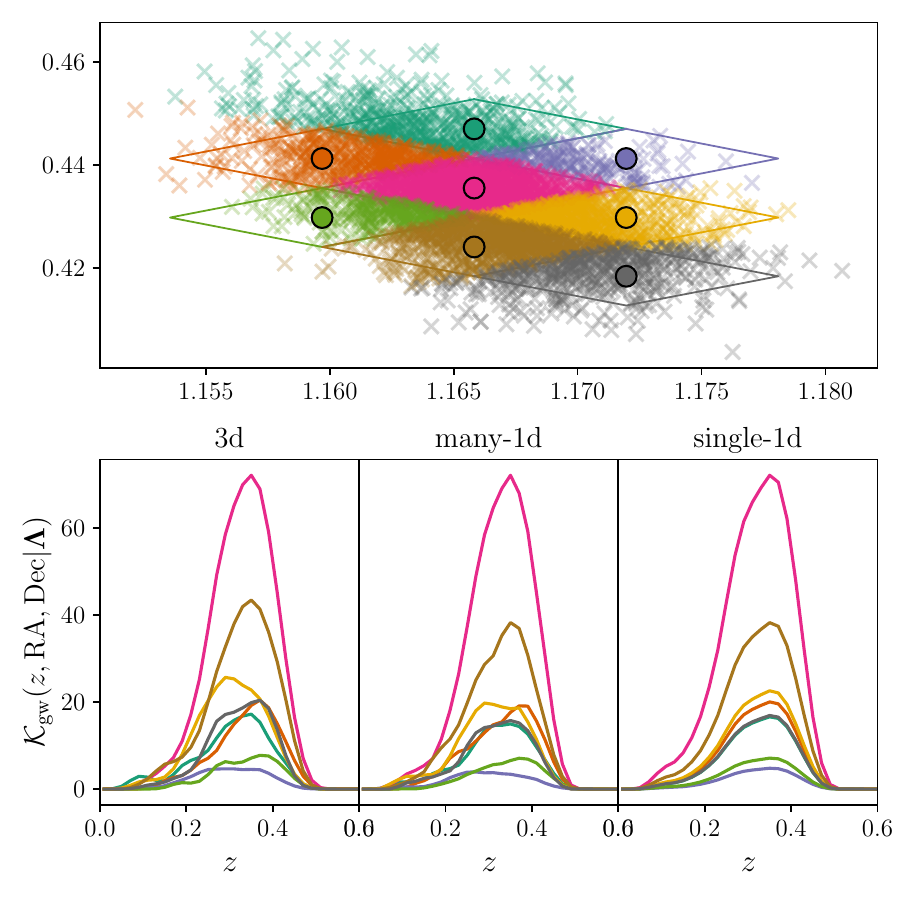}}
    \caption{Top: Pixelization of 90\% localization area of a mock GW event, with PE samples (marked with a cross) colored according to the pixel they fall into.
    Bottom: Different GW kernels implemented in \CHIMERAsecond. Each line represents the GW kernel in a pixel of the above mock GW event. Kernels are computed at single hyperparameter space points.}
    \label{fig:kde-comparison}
\end{figure}

\CHIMERAsecond includes three different KDE algorithms, which have been tested and validated against each other.
The first is the ``3D'' kernel implemented in the first release.
However, in \CHIMERAsecond, it is evaluated only on the portion of the redshift grid containing redshift samples.
Since the redshift grids must be pre-computed considering the cosmological priors, the redshift samples are contained in a smaller fraction - up to 1/3 - of the redshift grids at each MCMC step. 
In the remaining segments of the grids, the kernel is set to zero.

\medskip \noindent
In the ``many-1D'' kernel approach, the pixelization and the pre-computation of the redshift grids and $p_{\rm cat}$ remain the same as in the previous case. 
However, this algorithm approximates the GW kernel in each pixel of $\delta\Omega_i$, indexed with $k$, using 1D weighted KDEs for each pixel, for a total of $N^i_{\rm pix}$ KDEs.
Each KDE is trained using the $N^i_k$ PE samples that are comprised in the corresponding pixel, as illustrated in the top panel of \Cref{fig:kde-comparison}.
In other words, this algorithm computes the 1D redshift distributions for each pixel by marginalizing over the sky localization distribution.
The GW kernel within each pixel is then obtained by multiplying the 1D KDE by a pixel-dependent normalization factor given by the 2D KDE of $\{\hat{\Omega}\}_{j=1}^{N_{\rm s}}$ evaluated at the center of the pixel:
\begin{align}\label{eq:gw-kernel-many1d}
     \mathcal{K}_{\mathrm{gw}, i}(z,\hat{\Omega}_k \mid \boldsymbol{d}_i, \boldsymbol{\lambda}_c, \boldsymbol{\lambda}_{mg},\boldsymbol{\lambda}_m) \approx \nonumber \\ 
     \l. \text{KDE}\l[\l\{z^j_i \l| w^j_i = \f{p(m^j_{1,i}, m^j_{2,i} \mid \boldsymbol{\lambda}_m)}{\pi(m^{\dd,j}_{1,i}, m^{\dd,j}_{2,i}) \, \pi(d^j_{L,i})}\r\}_{j=1}^{N^i_k}\r. \r]\r|_{z} \times \nonumber \\ \l.\text{KDE} \l[\l\{\hat{\Omega}^j_i\r\}_{j=1}^{N_{\rm s}} \r]\r|_{\hat{\Omega}_k}.
\end{align}
The normalization factors can be pre-computed once, as they are independent of hyperparameters. 
Also in this case, at each MCMC step, the KDEs are evaluated only on the portions of the pre-computed redshift grids containing redshift samples. 
Two kernel options are available for the 1D KDEs: Gaussian and Epanechnikov, with the latter being slightly faster.
To ensure a consistent dataset dimension across pixels and events, the weighted PE samples in each pixel are binned. 
This enables vectorized computation over both pixels and events, eliminating the costly Python loops used in the previous algorithm. 

\medskip \noindent
The ``single-1D'' kernel algorithm is very similar to the "many-1D" approach, but only requires a single KDE computation per event rather than $N^i_{\rm pix}$ KDEs. 
Instead of splitting the samples by pixel, all $N_{\rm s}$ redshift samples are used to train a weighted 1D KDE, which represents the redshift distribution of the GW event weighted by the mass distribution. 
The GW kernel in each pixel is then estimated by multiplying this global weighted redshift distribution by the same pixel-dependent normalization factors:
\begin{align}\label{eq:gw-kernel-single1d}
     \mathcal{K}_{\mathrm{gw}, i}(z,\hat{\Omega}_k \mid \boldsymbol{d}_i, \boldsymbol{\lambda}_c, \boldsymbol{\lambda}_{mg}, \boldsymbol{\lambda}_m) \approx \nonumber \\ 
     \l. \text{KDE}\l[\l\{z^j_i \l| w^j_i = \f{p(m^j_{1,i}, m^j_{2,i} \mid \boldsymbol{\lambda}_m)}{\pi(m^{\dd,j}_{1,i}, m^{\dd,j}_{2,i}) \, \pi(d^j_{L,i})}\r\}_{j=1}^{N_{\rm s}}\r. \r]\r|_{z} \times \nonumber \\ \l.\text{KDE} \l[\l\{\hat{\Omega}^j_i\r\}_{j=1}^{N_{\rm s}} \r]\r|_{\hat{\Omega}_k}.
\end{align}
This algorithm assumes independence between the sky position and redshift.
Thus, this approximation is suitable when these two variables are weakly correlated.
As in the "many-1D" case, this KDE supports Gaussian or Epanechnikov kernels, its evaluation is restricted to the relevant portion of the pre-computed redshift grid, and the dataset can be binned.
Also, this algorithm is fully vectorized over the events.

\begin{figure}
    \centering
    \resizebox{\hsize}{!}{\includegraphics{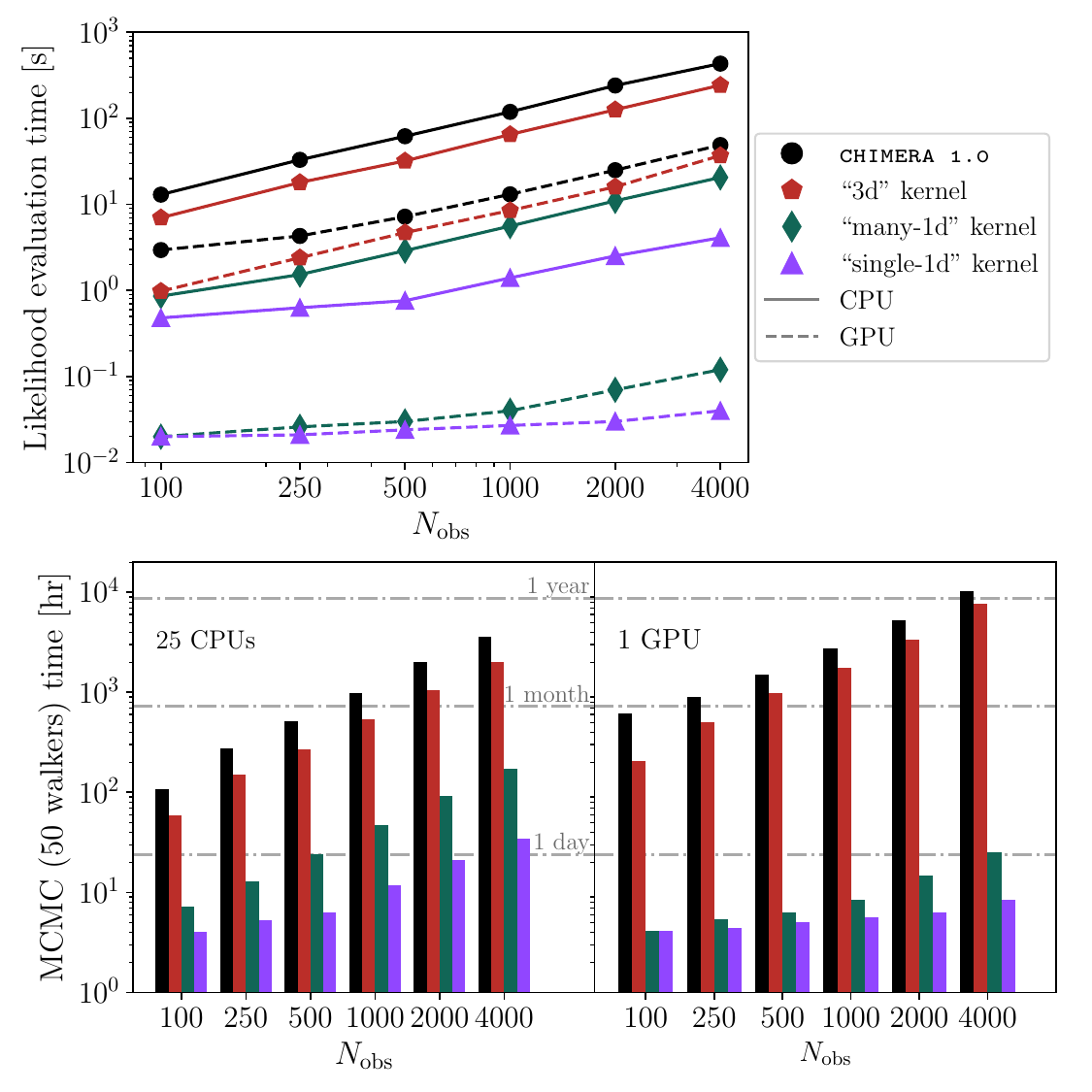}}
    \caption{Top: Likelihood evaluation times for catalogs with different numbers of events (each with 5000 PE samples) at a single hyperparameter space point, compared across different kernel algorithms. Solid curves show times on four Intel CPUs (2.0 GHz clock frequency), while dashed curves represent times on a single Nvidia A100 GPU. 
    Bottom: Run times for a full affine-invariant MCMC fit, using the \texttt{emcee} sampler with 50 walkers, on 25 CPUs (left panel) or one GPU (right panel).}
    \label{fig:kde-time-scaling}
\end{figure}

\subsection{Scaling performances}

The bottom panels of \Cref{fig:kde-comparison} compare the three different GW kernels considering a mock GW event with the localization area divided into 8 pixels.
The "3D" kernel distributions show fluctuations in less populated pixels. 
In contrast, the "single-1D" kernel smooths out these substructures, while the "many-1D" one provides only partial smoothing.
In \Cref{app:kernel-validation} we compare and validate the "single-1D" and "many-1D" against the implementation of \CHIMERAfirst.
In particular, \Cref{fig:corner-validation} highlights that the three kernels produce results that are consistent with each other, with no significant differences observed.

The new kernels yield the likelihood evaluation times shown in the top panel of \Cref{fig:kde-time-scaling}. 
Since the "3D" kernel is only calculated on a limited portion of the pre-computed redshift grid, the dependence of $t_{\rm KDE}$ on $N_{\rm z}$ is reduced, achieving a $2-3$ speed improvement over \CHIMERAfirst.
However, this algorithm still relies on a computationally expensive loop over the events, resulting in a linear increase of $t_{\rm KDE}$ with $N_{\rm obs}$.
The vectorized "many-1D" and "single-1D" algorithms mitigate this linear scaling, particularly on a GPU. 
The computational time for the "many-1D" kernel is further reduced compared to the "3D" approach due to the dataset's lower dimensionality, resulting in a $10-20\,(100-200)$ speed improvement over \CHIMERAfirst on the CPU (GPU), depending on the number of events.
The "single-1D" case, in turn, does not depend on the number of pixels and benefits from binning to reduce the dependence of $t_{\rm KDE}$ on $N_{\rm s}$, making it the fastest algorithm among the three.
This algorithm is $20-100\,(100-1000)$ times faster than the first release on the CPU (GPU), depending on the number of events.

The "many-1D" kernel reduces the time spent on KDE analysis from $80\%$ (as in \CHIMERAfirst) to $35\%$ for a fit of a catalog of 100 GW events with 5000 PE samples and $2 \times 10^7$ injections.
The "single-1D" algorithm further cuts this fraction to $15\%$.
With the "many-1D" algorithm, the GW kernel \eqref{eq:gw-kernel} and bias term \eqref{eq:bias} evaluations take a similar time, but for larger datasets, the kernel becomes again the main bottleneck. 
In contrast, in the "single-1D" case, the total time is dominated by the bias term, which remains the limiting factor even with more events — since the required injections for numerical accuracy is expected to scale as $\propto N^2_{obs}$ \citep{Talbot:2023pex}.

The bottom panels of \Cref{fig:kde-time-scaling} show the total run time of a full affine-invariant MCMC fit using the \texttt{emcee} sampler with 50 walkers. 
In the left panel, run times are shown for 25 CPUs, the maximum number that can be used to parallelize walker moves.
In the right panel, run times on a single GPU are presented, where walkers evolve sequentially. 
Poorly vectorized algorithms, such as \CHIMERAfirst and the "3D" kernel, are less efficient on GPUs than on multiple CPUs, despite faster single-point evaluation on GPUs.
In contrast, the "many-1D" and "single-1D" kernels perform better on GPUs and exhibit a weaker linear dependence on the number of events.
These KDE algorithms are a key advancement, preparing \CHIMERAsecond for next-generation detectors that will detect 1000 times more events than current observatories.
Finally, we stress that for less parallelizable methods, such as standard MCMC, parallel tempering MCMC, Hamiltonian MCMC, or nested-sampling methods - where fewer chains are evolved in parallel compared to affine-invariant MCMC options - the advantage of a single GPU over multiple CPUs becomes even more significant.

Although the "single-1D" kernel is the most efficient and produces results consistent with the other algorithms, \CHIMERAsecond retains all three KDE methods.
This ensures flexibility, allowing the code to handle also real GW events with complex posterior distributions, where the assumption behind the "single-1D" kernel may not hold.

\section{Data}\label{sec:data}

In the following, we outline the process used to generate the three mock GW catalogs for the MCMC analyses, following a method similar to that of \cite{Borghi:2023opd}.

\subsection{GW populations}\label{eq:subsec-gw-samples}

The mock GW catalogs were generated starting from a mock galaxy catalog and populating galaxies with CBC events drawn from a fiducial population model. These are detailed in the following paragraphs.

\begin{figure}
    \centering
    \resizebox{\hsize}{!}{\includegraphics{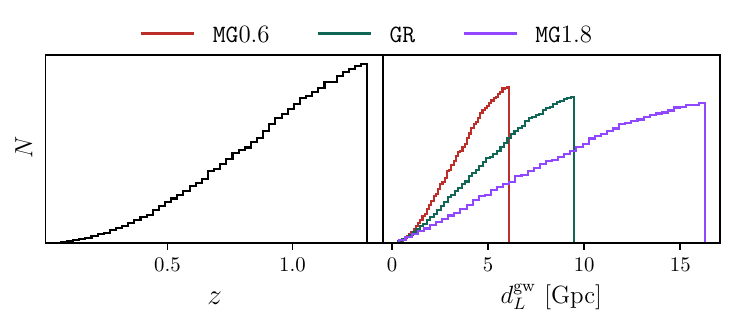}}
    \caption{Histograms of galaxy-catalog redshifts (right) and of corresponding luminosity distances in the three MG scenarios considered in this work (left). $\texttt{MG}0.6$ is the sample with $\Xi_0 = 0.6$, $\texttt{GR}$ with $\Xi_0 = 1$, and $\texttt{MG}1.8$ with $\Xi_0 = 1.8$}
    \label{fig:z-parent-sample}
\end{figure}

\medskip \noindent
Similarly to \cite{Borghi:2023opd}, the galaxy catalog considered in this work is a complete subsample from the MICE Grand Challenge light-cone simulation (v2) \citep{Carretero:2014ltj, Fosalba:2013wxa, Fosalba:2013mra, Hoffmann:2014ida}.
The MICEv2 catalog covers one octant of the sky and is designed to mimic a complete DES-like survey up to an observed magnitude of $i<24$ at redshift $z<1.4$.
The galaxy catalog was obtained by considering only galaxies with stellar masses $\log M_* / \msun > 10.5$ and with a uniform comoving volume redshift distribution up to $z<1.3$, resulting in $1.6$ million galaxies.
This cut is aligned with the assumption that the binary merger rate follows the stellar mass distribution.
This assumption is widely adopted in the current literature through absolute magnitude cuts and luminosity weighting \citep{LIGOScientific:2018gmd, Finke:2021aom, Gray:2021sew, LIGOScientific:2021aug, Muttoni:2023prw, Alfradique:2025tbj}.
In the left panel of \Cref{fig:z-parent-sample}, we show the redshift distribution of the galaxy catalog.

\medskip \noindent
The sample of GW events was created by populating the galaxy catalog with CBC events.
The sky position and redshift of each event are determined by the host galaxy, while the luminosity distance was obtained by assuming fiducial cosmological, and MG propagation models. 
We chose the same cosmological model adopted in the MICE simulation: a flat $\Lambda$CDM universe with $H_0 = \SI{70}{\kilo\meter\per\second\per\mega\parsec}$, $\Omega_{m,0} = 0.25$.
We parameterized MG propagation with two parameters ($\Xi_0, n$) as in \Cref{eq:dL-mg}.
Binary masses in the source frame were drawn from a mass distribution factorized as
\begin{align}
    p(m_1,m_2\mid\boldsymbol{\lambda}_m) = p(m_1\mid\boldsymbol{\lambda}_m) p(m_2\mid m_1, \boldsymbol{\lambda}_m).
\end{align}
The primary mass follows a power law with spectral index $-\alpha$, truncated in the $[m_{\rm low},m_{\rm high}]$ range, with an additional Gaussian peak centered at $\mu_g$ with a width of $\sigma_g$, weighted by $\lambda_g$.
The lower edge of the power law is smoothed using a parameter $\delta_m$.
The secondary mass follows a smoothed power law with spectral index $\beta$, truncated in the $[m_{\rm low}, m_1]$ range.
The chosen mass model, fully described in Appendix B of \cite{LIGOScientific:2020kqk}, is consistent with the latest GWTC-3 constraints \citep{KAGRA:2021duu}.
For the merger-rate evolution, we used the Madau-Dickinson parameterization~\citep{Madau:2014bja, Fishbach:2017zga}:
\begin{equation}
    \psi(z ; \boldsymbol{\lambda}_r) \propto \f{(1+z)^\gamma}{1+\l(\f{1+z}{1+z_p}\r)^{\gamma+\kappa}}.
\end{equation}

For the cosmological, rate, and mass hyperparameters a single fiducial value was chosen.
For the MG hyperparameters, three combinations are explored, resulting in three different GW populations.
One case, $\Xi_0 = 1$, corresponds to GR with no modified GW propagation.
The other cases, $\Xi_0 = 0.6$ and $\Xi_0 = 1.8$, are at the boundaries of the 1-$\sigma$ constraints obtained by \cite{Mancarella:2021ecn} using O3 data and considering only features of the population models.
The value of $n$ is set to 2.7 in all scenarios.
Starting from the same redshift distribution, the corresponding luminosity distance distributions of the three populations are expected to span different ranges due to the different values of $\Xi_0$, affecting both the number and the distributions of detected events.
These differences can be inspected in the right panel of \Cref{fig:z-parent-sample}, where we show the luminosity distance distributions of the galaxy catalog in the three different MG scenarios. 

\Cref{tab:hyperparams_summary} summarizes the hyperparameters describing the GW populations, their fiducial values, and the prior used in the following MCMC analyses.

\begin{table}[t]
\caption{Summary of hyperparameters and priors adopted.}
\centering\resizebox{\hsize}{!}{
\begin{tabular}{llcc}
\hline
\hline
Symbol & Description & Fiducial Value & Prior \\
\hline
\multicolumn{4}{l}{\textbf{Cosmology (flat $\Lambda$CDM)}} \\
$H_0$ & Hubble constant [km/s/Mpc] & 70.0 & $\mathcal{U}(10.0, 200.0)$ \\
$\Omega_{\rm m,0}$ & Matter energy density & 0.25 & Fixed \\
\hline
\multicolumn{4}{l}{\textbf{Modified gravitational wave propagation}} \\
$\Xi_0$ & See \Cref{eq:dL-mg} & 0.6, 1, 1.8 & $\mathcal{U}(0.1, 10)$ \\
$n$ & See \Cref{eq:dL-mg} & 2.7 & $\mathcal{U}(1, 5)$  \\
\hline
\multicolumn{4}{l}{\textbf{Mass distribution (Power Law + Gaussian Peak)}} \\
$\alpha$ & Primary power law slope & 3.4 & $\mathcal{U}(1.5, 12)$ \\
$\beta$ & Secondary power law slope & 1.1 & $\mathcal{U}(-4, 12)$ \\
$\delta_m$ & Smoothing parameter $[\msun]$ & 4.8 & $\mathcal{U}(0.01, 10.0)$ \\
$m_{\rm high}$ & Power laws upper limit $[\msun]$ & 87.0 & $\mathcal{U}(50, 200)$ \\
$\mu_{\rm g}$ & Gaussian peak position $[\msun]$ & 34.0 & $\mathcal{U}(2, 50)$ \\
$\sigma_{\rm g}$ & Gaussian peak width $[\msun]$ & 3.6 & $\mathcal{U}(0.4, 10)$ \\
$\lambda_{\rm g}$ & Gaussian peak weight & 0.039 & $\mathcal{U}(0.01, 0.99)$ \\
\hline
\multicolumn{4}{l}{\textbf{Rate evolution (Madau-like)}} \\
$\gamma$ & Slope at $z<z_p$ & 2.7 & $\mathcal{U}(0, 12)$ \\
$\kappa$ & Slope at $z>z_p$ & 3.0 & $\mathcal{U}(0, 6)$ \\
$z_{\rm p}$ & Peak redshift & 2.0 & $\mathcal{U}(0, 4)$ \\
\hline
\end{tabular}}
\tablefoot{The symbol $\mathcal{U}(\cdot)$ denotes a uniform prior distribution.}
\label{tab:hyperparams_summary}
\end{table}

\begin{figure*}[ht]
    \centering
    \includegraphics[width=\textwidth]{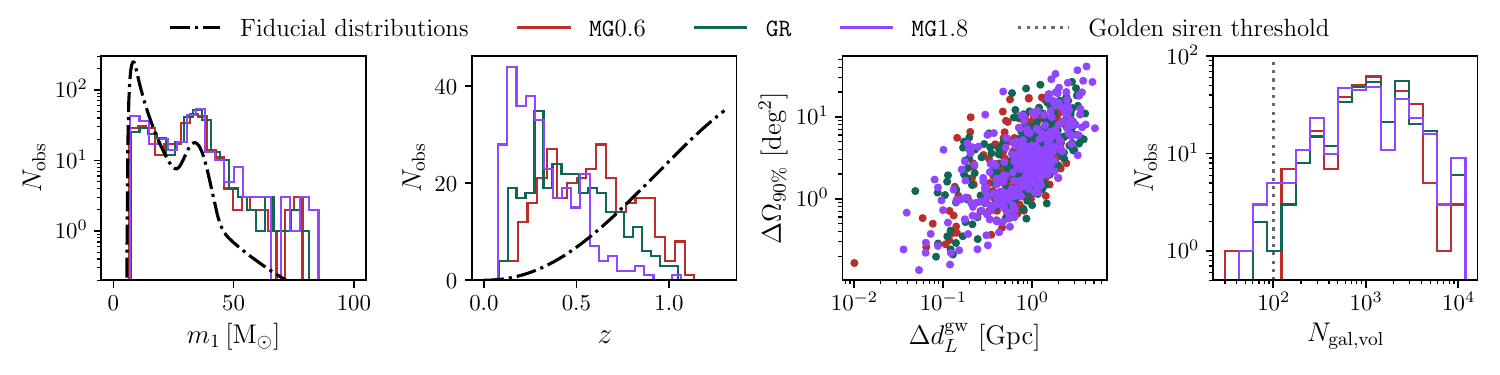}
    \caption{Properties of three GW mock catalogs. From left to right: primary mass distributions, redshift distributions, luminosity distance, localization area uncertainties, and histograms of the number of galaxies in the localization volume.}
    \label{fig:cat-props}
\end{figure*}

\subsection{GW detection and PE generation}
The sample of GW events was analyzed using \texttt{GWFAST} \citep{Iacovelli:2022bbs, Iacovelli:2022mbg} to identify detectable events.
We assume the CBC event to be quasi-circular, non-precessing BBH system.
Each CBC waveform is characterized by 15 detector-frame parameters:
\begin{equation}
    \boldsymbol{\theta}^\dd = \{\mathcal{M}_c, \eta, d_L, \theta, \phi, \iota, \chi^z_1, \chi^z_2, \phi, t_c, \Phi_c \},
\end{equation}
where $\mathcal{M}_c$ is the redshifted chirp mass, $\eta$ is the symmetric mass ratio, $d_L$ is the binary luminosity distance, $(\theta,\phi)$ are the sky position angles, $\iota$ is the inclination angle between the orbital angular momentum and the line of sight, $\chi^z_{1,2}$ are the spin projections along the orbital angular momentum, $\psi$ is the polarization angle, $t_c$ is the coalescence time, and $\Phi_c$ is the phase at coalescence.
While the first five parameters were generated based on the galaxy catalog and the population models as explained in the previous subsection, the remaining parameters were drawn from specific distributions: the spin components are uniformly distributed in $[-1,1]$, the inclination angle is uniform in $\cos\iota$ over $[0,\pi]$, the polarization angle and coalescence phase are uniformly distributed in $[0,\pi]$ and $[0,2\pi]$, respectively, and the coalescence time - expressed in units of fraction of a day - is uniform in $[0,1]$.

The waveform signal, simulated using the \texttt{IMRPhenomHM} \citep{London:2017bcn} approximant, is injected into a noise realization of each detector.
We considered a network configuration that includes the two LIGO interferometers in the USA \citep{LIGOScientific:2014pky}, the Virgo interferometer in Italy \citep{VIRGO:2014yos}, the KAGRA interferometer in Japan \citep{Aso:2013eba}, and the planned LIGO interferometer in India \citep{LIGO:M1100296}.
We assumed the publicly available\footnote{We used \texttt{AplusDesign} for the three LIGO detectors, \texttt{avirgo O5low NEW} for Virgo, and \texttt{kagra 80Mpc} for KAGRA. These curves are available at \url{https://dcc.ligo.org/LIGO-T2000012/public}.} sensitivity curves representative of the O5 observing run \citep{KAGRA:2013rdx}, with 100\% duty cycle.
We then analyzed the injected signals with \texttt{GWFAST}, estimating the network's match-filtered signal-to-noise ratio (S/N) and computing the Fisher-information-matrix (FIM) and its inverse.
The individual GW event likelihood for the detector-frame parameters is approximated as a multivariate Gaussian, with the covariance matrix given by the inverse FIM.
In \cite{Borghi:2023opd}, we checked that the Fisher matrix approach was a good approximation for high S/N events, such as the ones considered in the following analyses.
The PE samples for $\boldsymbol{\theta}^\dd$ are drawn from this approximated likelihood using the \texttt{emcee} sampler.
The priors used are uniform in the $[0,10^5]\msun$, $[0,1/4]$ range, and in $[0,10^5]\si{\giga\parsec}$ for $\mathcal{M}_c$, $\eta$, and $d_L$, respectively, while all other waveform parameters were bounded in the same physical ranges from which they were drawn.
The prior also includes the Jacobian of the transformation $(\mathcal{M}_c, \eta) \to (m^\dd_1, m^\dd_2)$ to ensure that the binary masses PE samples are uniformly distributed.

To create the injection set, we adopted a similar process, excluding the unnecessary PE generation step.
We used the same $N_{\rm inj} = 2 \times 10^7$ injections as in the O5-like mock catalog described in \cite{Borghi:2023opd}.

\begin{figure*}[ht]
    \centering
    \resizebox{0.49\hsize}{!}{\includegraphics{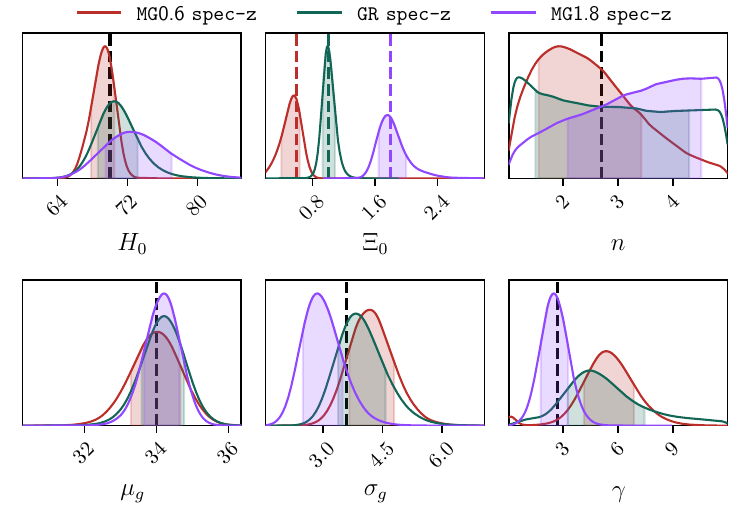}}
    \resizebox{0.49\hsize}{!}{\includegraphics{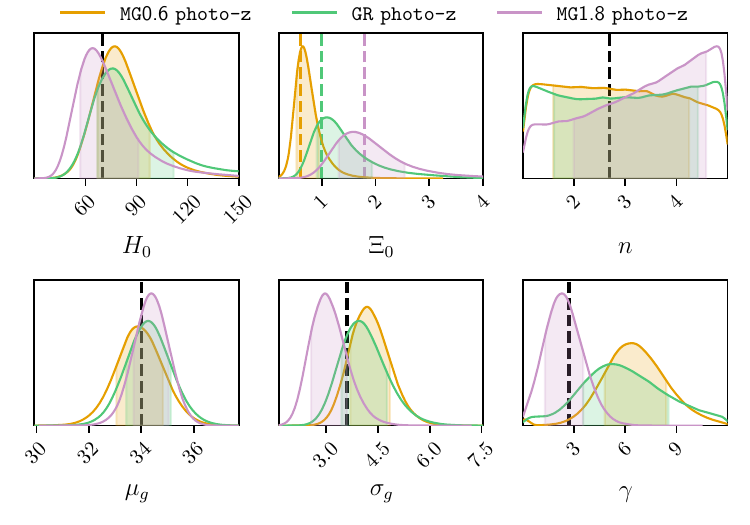}}
    \vspace*{-1em}
    \caption{Marginalized posterior distributions for selected hyperparameters. These constraints are obtained in the first MCMC configuration (inference on all hyperparameters) and assuming a spectroscopic (left panel) or photometric (right panel) galaxy catalog. The colored areas under each posterior represent the 68\% C.L. The dotted lines indicate the hyperparameter fiducial values.}
    \label{fig:1d-allFree-spec-photo}
\end{figure*}

\subsection{GW catalogs selection and properties}
Based on the population models described in \Cref{eq:subsec-gw-samples} and assuming a local BBH merger rate density of $R_0 = \SI{17}{\per\cubic\giga\parsec\per\years}$ \citep{KAGRA:2021duu}, the expected number of events per year out to $z = 1.3$ (the upper limit of our galaxy catalog) is:
\begin{equation}
    \f{\dd N_{\rm cbc}}{\dd t_\dd} = R_0 \int_0^{1.3} \dd z \f{\psi(z ; \boldsymbol{\lambda}_r)}{1+z} \f{\dd V_c}{\dd z}(z; \boldsymbol{\lambda}_c) \approx 1.4 \times 10^{4}\,\si{\per\years} .
\end{equation}
We note that this rate is strongly affected by the chosen value of $R_0$, which is still largely unbounded from O3 data $R_0 = 17^{10}_{-6.7} \si{\per\cubic\giga\parsec\per\years}$.
From this total, the number of detected events with S/N>8 is approximately 6200, 3100, and 1100 in the \texttt{MG}0.6, \texttt{GR}, and \texttt{MG}1.8 scenarios, respectively.
This difference can be understood from the luminosity distance distributions of the three GW samples in \Cref{fig:z-parent-sample}, as higher (lower) $\Xi_0$ values map the true redshifts into higher (lower) luminosity distances, affecting the detection rates.
Similarly, the number of events with S/N>20 per year is approximately 915, 340, and 100 in the three scenarios. 

We selected 300 events at S/N>20 from each GW catalog.
This corresponds to roughly one year of observation assuming that GR holds, or three years (4.5 months) for $\Xi_0 = 1.8$ ($\Xi_0 = 0.6$). 
Keeping the same number of events enables a fair comparison of the derived constraints, which strongly depend on the number of events. 
On the other hand, since different values of $\Xi_0$ have an impact on the number of GW events detected per year, we also analyzed results within a fixed time frame using two additional catalogs: 900 sources for \texttt{MG}0.6 and 100 for \texttt{MG}1.8. 
In \Cref{sec:results}, we discuss the impact on the results of this different scenario.

The properties of the three catalogs are illustrated in \Cref{fig:cat-props}.
While the primary mass distribution appears similar across the three cases, the redshift histogram of the \texttt{MG}1.8 catalog shows more events at lower redshifts. 
Again, this effect can be understood since high-redshift GW events are less likely to be detected as they correspond to very high-luminosity distances in this scenario.
For the opposite reason, the \texttt{MG}0.6 catalog extends to higher redshifts.
Since the cut in S/N is the same for all three catalogs, the luminosity distance and localization area measurement uncertainties are similar.
The events in the \texttt{MG}1.8 catalog typically have fewer galaxies in the localization volume, which is expected due to the lower number of galaxies at lower redshifts.
In particular, this catalog contains seven ``golden sirens'', defined here as BBHs with 100 or fewer galaxies within their localization volume.
In comparison, the \texttt{MG}0.6 and \texttt{GR} catalogs have one and three golden sirens, respectively, since their redshift distributions are shifted to higher values.

\begin{table*}[ht]
    \caption{Constraints on $H_0$ and MG hyperparameters.}
    \centering
    \begin{tabular}{r|c@{\hspace{3pt}}c c@{\hspace{3pt}}c c@{\hspace{3pt}}c|c@{\hspace{3pt}}c c@{\hspace{3pt}}c c@{\hspace{3pt}}c}
        \hline\\[-1.5ex]
        \multirow{2}{*}{Catalog} & \multicolumn{6}{c|}{\texttt{spec-z}} & \multicolumn{6}{c}{\texttt{photo-z}} \\
        \cline{2-13}
        & \multicolumn{2}{c}{$H_0$} & \multicolumn{2}{c}{$\Xi_0$} & \multicolumn{2}{c|}{$n$} & \multicolumn{2}{c}{$H_0$} & \multicolumn{2}{c}{$\Xi_0$} & \multicolumn{2}{c}{$n$} \\
        \hline\\[-1.5ex]
        \multirow{3}{*}{$\text{MG}0.6$} 
        & $70.11^{+0.61}_{-0.61}$ & (0.9\%) & \multicolumn{2}{c}{$\cdot$} & \multicolumn{2}{c|}{$\cdot$} & $71.9^{+3.8}_{-3.7}$ & (5.2\%) & \multicolumn{2}{c}{$\cdot$} & \multicolumn{2}{c}{$\cdot$} \\[1pt]
        & \multicolumn{2}{c}{$\cdot$} & $0.56^{+0.11}_{-0.13}$ & (22\%) & $2.42^{+0.79}_{-0.68}$ & (30\%) & \multicolumn{2}{c}{$\cdot$} & $0.60^{+0.15}_{-0.14}$ & (24\%) & $1.46^{+0.81}_{-0.34}$ & (39\%) \\[1pt]
        & $69.3^{+1.2}_{-1.4}$ & (1.9\%) & $0.54^{+0.10}_{-0.14}$ & (22\%) & $2.35^{+1.08}_{-0.80}$ & (40\%) & $80^{+18}_{-13}$ & (19\%) & $0.70^{+0.25}_{-0.17}$ & (30\%) & $2.9^{+1.4}_{-1.3}$ & (47\%) \\[2pt]
        \hline\\[-2ex]
        \multirow{3}{*}{GR} 
        & $70.45^{+0.59}_{-0.56}$ & (0.82\%) & \multicolumn{2}{c}{$\cdot$} & \multicolumn{2}{c|}{$\cdot$} & $71.4^{+4.4}_{-4.3}$ & (6.1\%) & \multicolumn{2}{c}{$\cdot$} & \multicolumn{2}{c}{$\cdot$} \\[1pt]
        & \multicolumn{2}{c}{$\cdot$} & $0.99^{+0.08}_{-0.08}$ & (8\%) & $2.5^{+1.6}_{-1.1}$ & (54\%) & \multicolumn{2}{c}{$\cdot$} & $1.00^{+0.12}_{-0.10}$ & (11\%) & $2.6^{+1.6}_{-1.2}$ & (54\%) \\[1pt]
        & $70.7^{+2.4}_{-2.1}$ & (3.2\%) & $1.00^{+0.08}_{-0.07}$ & (7.5\%) & $2.8^{+1.5}_{-1.3}$ & (50\%) & $82^{+30}_{-15}$ & (27\%) & $1.27^{+0.67}_{-0.36}$ & (41\%) & $3.1 \pm 1.4$ & (45\%) \\[2pt]
        \hline\\[-2ex]
        \multirow{3}{*}{$\text{MG}1.8$} 
        & $70.4^{+0.5}_{-0.5}$ & (0.72\%) & \multicolumn{2}{c}{$\cdot$} & \multicolumn{2}{c|}{$\cdot$} & $72.3^{+4.9}_{-4.6}$ & (6.5\%) & \multicolumn{2}{c}{$\cdot$} & \multicolumn{2}{c}{$\cdot$} \\[1pt]
        & \multicolumn{2}{c}{$\cdot$} & $1.79^{+0.36}_{-0.19}$ & (15\%) & $2.37^{+0.99}_{-0.82}$ & (38\%) & \multicolumn{2}{c}{$\cdot$} & $1.73^{+0.38}_{-0.23}$ & (18\%) & $2.5^{+1.5}_{-1.1}$ & (52\%) \\[1pt]
        & $73.0^{+4.1}_{-3.5}$ & (5.2\%) & $1.79^{+0.20}_{-0.15}$ & (10\%) & $3.5^{+1.1}_{-1.4}$ & (36\%) & $70^{+21}_{-13}$ & (24\%) & $1.78^{+0.76}_{-0.47}$ & (35\%) & $3.5^{+1.1}_{-1.5}$ & (37\%) \\[2pt]
        \hline
    \end{tabular}
    \tablefoot{The central values are the median of the 1D marginalized distributions. Errors are reported as the 68\% C.L. around the median. The percentage error shown is the mean between the upper and lower percentage errors. For each catalog: the first raw shows results when inferring only $H_0$ and population hyperparameters, keeping MG hyperparameters fixed; the second raw infers MG and population hyperparameters only, keeping $H_0$ fixed; the last row infers all hyperparameters.}
    \label{tab:res-H0-MG}
\end{table*}

\section{Results}\label{sec:results}

In this section, we discuss the results obtained.
For each GW catalog, we considered both spectroscopic (\texttt{spec-z}) and photometric uncertainties on galaxy redshifts:
\begin{equation}
\tilde\sigma_{z,g} = \begin{cases} 
0.001(1+z), & \texttt{spec-z}; \\ 
0.05(1+z), & \texttt{photo-z}.
\end{cases}
\end{equation}
Spectroscopic galaxy catalogs can be obtained by expanding the currently available catalog (GLADE+ \citealt{Dalya:2021ewn}) with future data.
For example, the all-sky ESA Euclid survey \citep{EUCLID:2011zbd} will measure spectroscopic redshift in the $0.9<z<1.8$ range with an accuracy level of $\tilde\sigma_{z,g} /(1+z) \lesssim 0.001$; DESI \citep{DESI:2016fyo} was planned to observe about one-third of the sky and cover the redshift range of $0.4 < z < 2.1$; the WST will map roughly half of the sky up to a target redshift of 1.5 \citep{WST:2024rai}.
Photometric redshifts are also available with ongoing surveys.
For instance, the DES survey reached $\tilde\sigma_{z,g} \sim 0.01$ \citep{DES:2020ebm} over a smaller area, with the potential to improve to $\tilde\sigma_{z,g} \sim 0.007$ using advanced techniques \citep{DES:2019bxr}.
Euclid and the upcoming Rubin observatory \citep{LSST:2008ijt} are expected to provide photometric redshifts with an accuracy of $\tilde\sigma_{z,g} \sim 0.05(1+z)$ \citep{Euclid:2020gbk, Euclid:2022vkk}.

For each GW and galaxy catalog (with different redshift assumptions), we sampled the posterior with an MCMC approach, exploring the following configurations: 
\begin{enumerate}
    \item inference on all hyperparameters listed in \Cref{tab:hyperparams_summary};
    \item inference only on MG and population hyperparameters, fixing $H_0$ to its fiducial value;
    \item inference on $H_0$ and population hyperparameters, fixing MG ones to their fiducial values.
\end{enumerate}
In all cases, $\Omega_{m,0}$ is fixed.
For each GW catalog, this results in six MCMC fits for a total of 18 runs. 
The priors used are summarized in \Cref{tab:hyperparams_summary}.
We used the $\texttt{emcee}$ sampler with 100 walkers and evolved the chains until the number of samples was at least 50 times larger than the integrated autocorrelation time for all the hyperparameters.
The "many-1D" kernel is employed, enabling each fit to converge in about 18 hours on a single GPU.
This extensive series of 18 tests, which cumulatively represent a population of about 5000 events, would not have been possible with \CHIMERAfirst, which was already at its limit with only 100 sourced events.
Using \CHIMERAfirst for this work would have taken approximately 270 CPU days or 18 GPU months (see \Cref{fig:kde-time-scaling}).

\begin{figure}[ht]
    \centering
    \resizebox{\hsize}{!}{\includegraphics{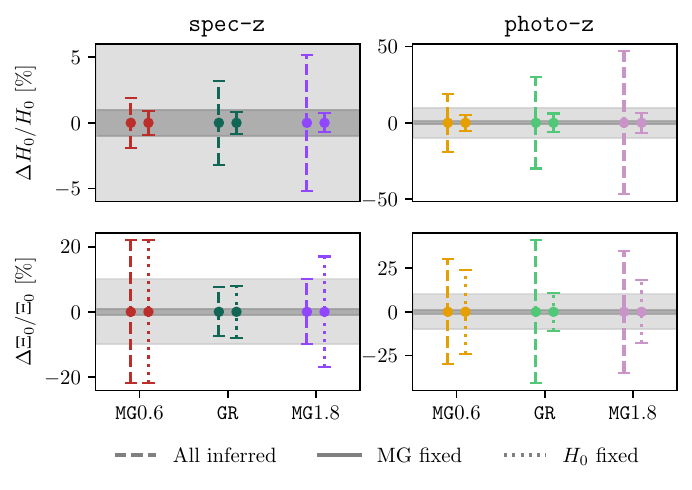}}
    \caption{Percentage precision on $H_0$ and $\Xi_0$ obtained with three GW catalogs across the various MCMC configurations. Dark (light) gray band represents the 1\% (10\%) error range.}
    \label{fig:H0Xi0-errors}
\end{figure}

\begin{figure}[ht]
    \centering
    \resizebox{\hsize}{!}{\includegraphics{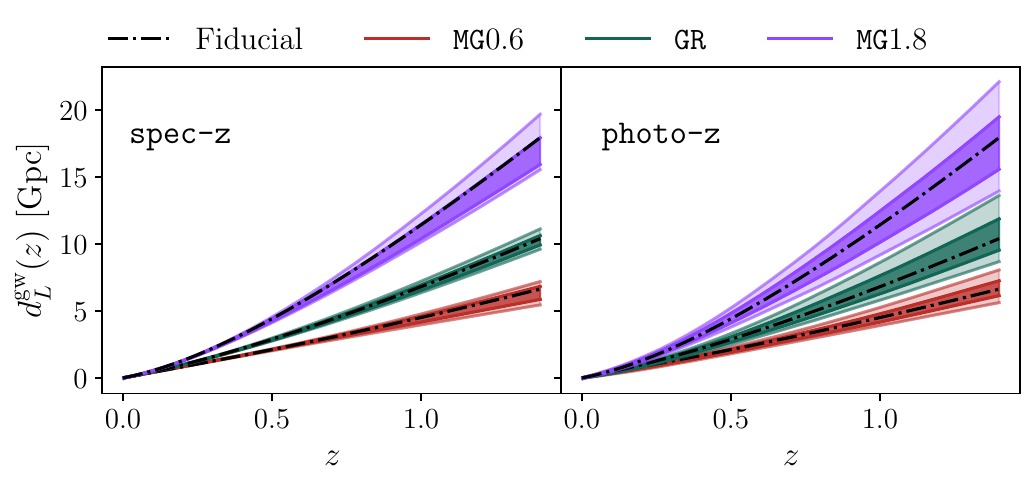}}
    \caption{Projected constraints on $d^{\mathrm{gw}}_L-z$ relation \eqref{eq:dL-mg} obtained in first MCMC configuration. The darker (lighter) contours represent the 68\% (95\%) C.L.}
    \label{fig:dL-z-constraints}
\end{figure}

\begin{figure*}[ht]
    \centering
    \resizebox{\hsize}{!}{\includegraphics{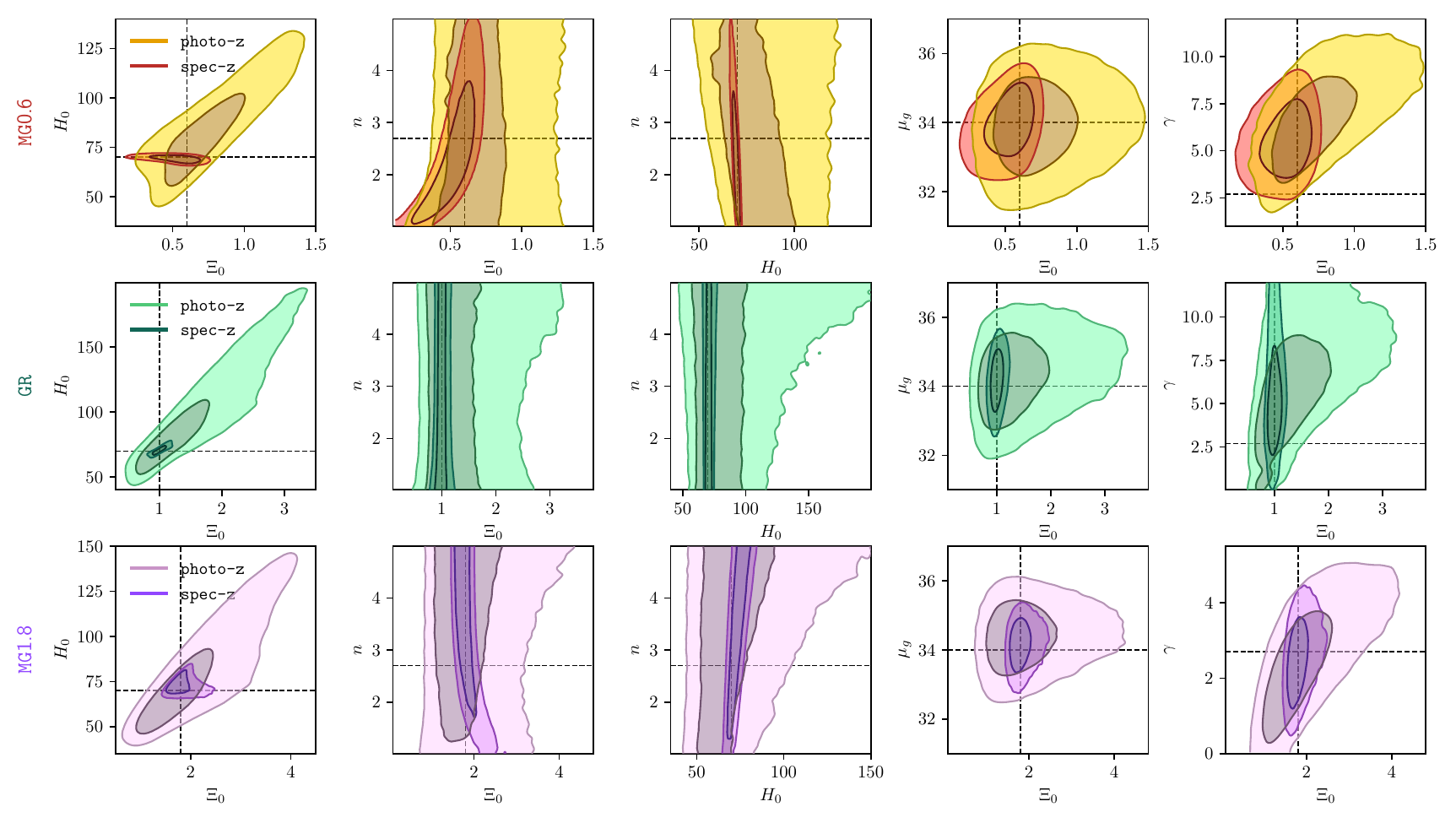}}
    \caption{2D contours between cosmological, MG, and selected population hyperparameters across the different GW and galaxy catalogs. The $68\%$ and $95\%$ confidence level regions are indicated by dark and light colors, respectively. The dotted lines represent the hyperparameter fiducial values.}
    \label{fig:table-2d-contours}
\end{figure*}

\subsection{Hyperparameter constraints}
The left panel of \Cref{fig:1d-allFree-spec-photo} shows the 1D marginalized posterior distributions obtained in the first MCMC configuration using spectroscopic galaxy redshifts for the three GW catalogs.
The constraints are shown for cosmological, MG, and selected population hyperparameters.
For the latter, we focused on the position and width of the Gaussian mass peak and the low-$z$ slope of the rate.
Indeed, these are the hyperparameters that mostly correlate with the cosmological and MG ones.
The right panel of \Cref{fig:1d-allFree-spec-photo} shows the corresponding results using photometric galaxy redshifts.
The colored areas represent the $68\%$ confidence levels (C.L.s) around the median of the distributions.
\Cref{tab:res-H0-MG} summarizes the median, the $68\%$ C.L., and the percentage precision for cosmological and MG hyperparameters across all MCMC configurations.
The percentage precisions on $H_0$ and $\Xi_0$ across the three catalogs and the different MCMC configurations are also shown in \Cref{fig:H0Xi0-errors}.
To visualize how constraints on MG and cosmological hyperparameters vary across the three GW catalogs and between photometric and spectroscopic redshifts, we project them on the luminosity distance-redshift relation \eqref{eq:dL-mg}, as shown in \Cref{fig:dL-z-constraints}.

\medskip \noindent
{\em \emph{We now discuss the results with spectroscopic galaxy redshifts}}.
$H_0$ and $\Xi_0$ were recovered without bias at 68\% C.L. across all MCMC configurations.
In the second MCMC configuration, $\Xi_0$ was constrained with a precision of $22\%$, $7.5\%$, and $10\%$ for $\Xi_0 = 0.6$, $1$, and $1.8$, respectively.
This allows the three scenarios to be distinguished at $68\%$ C.L., as shown in the right panel of \Cref{fig:1d-allFree-spec-photo}.
Notably, fixing $H_0$ does not improve constraints on the MG hyperparameters (see \Cref{tab:res-H0-MG}).

The precision on $H_0$ depends on whether MG hyperparameters are marginalized or fixed.
When the MG hyperparameters were fixed to their fiducial values, $H_0$ was recovered with a precision on the order of or slightly better than $1\%$. 
The best constraint, $0.72\%$, was obtained in the \texttt{MG}1.8 case, likely due to its higher number of golden sirens compared to the other catalogs.
In fact, when the MG hyperparameters are not inferred, the precision on $H_0$ improves with the number of golden sirens in the catalog.
However, when MG hyperparameters are marginalized, the precision on $H_0$ degrades by factors of $\sim2.1$, $\sim3.9$, and $\sim7.2$ for the \texttt{MG}0.6, \texttt{GR}, and \texttt{MG}1.8 cases, respectively. 
In this scenario, the best result is found with the \texttt{MG}0.6 catalog, suggesting that the $H_0$ precision is no longer strongly influenced by the number of golden sirens.

The posterior distributions for $n$ remain nearly flat, even when $H_0$ is not marginalized.

\medskip \noindent
\emph{{\em With regard to the results with photometric galaxy redshifts},}
switching from spectroscopic to photometric redshifts does not introduce biases, but significantly weakens the constraints on $H_0$ and $\Xi_0$.
Specifically, the precision on $\Xi_0$ degrades by a factor of $\sim 1.4$, $\sim5.4$, and $\sim 3.5$ for the \texttt{MG}0.6, \texttt{GR}, and \texttt{MG}1.8 cases, respectively. 
Fixing $H_0$ mitigates this degradation, and $\Xi_0$ is recovered with a precision that is on average only $1.2$ times worse than the spectroscopic case with fixed $H_0$.
The precision on $H_0$ degraded by factors of $\sim 5 - 7$ when the MG hyperparameters were fixed, and by $\sim 4-10$ when $\Xi_0$ and $n$ were marginalized.
As in the spectroscopic case, $n$ remains unconstrained in the photometric scenario.

\medskip \noindent
{\em \emph{Concerning constraints for a fixed time frame}}, we tested how constraints change when considering a fixed time frame of a one-year of observation. To do this, we used the additional catalogs mentioned above of 900 and 100 sources for the \texttt{MG}1.8 and \texttt{MG}0.6 cases. We performed a test considering spectroscopic galaxies and inferring both MG and cosmological hyperparameters. 
We find that in the \texttt{MG}0.6 case, the constraint on $\Xi_0$ improves from $22\%$ to $17\%$, while the precision on $H_0$ remains approximately the same.
In the \texttt{MG}1.8 case, instead, the lower number of events per year degrades the precision of $\Xi_0$ from $10\%$ to $14\%$ and of $H_0$ from $5.2\%$ to $8.5\%$.

\medskip \noindent
{\em \emph{With regard to population hyperparameters}}, the Gaussian peak position and width of the primary mass distribution are recovered within the 68\% CL across all combinations of GW catalogs and redshift error assumptions.
The redshift error choice has only a marginal effect on the posteriors, and fixing either MG or cosmological hyperparameters has no impact on these hyperparameters.
This implies that population hyperparameter constraints are mainly driven by the GW data rather than the galaxy catalog used.

\begin{figure*}[ht]
\centering
\begin{minipage}{0.33\textwidth}
    \centering\includegraphics[width=\textwidth]{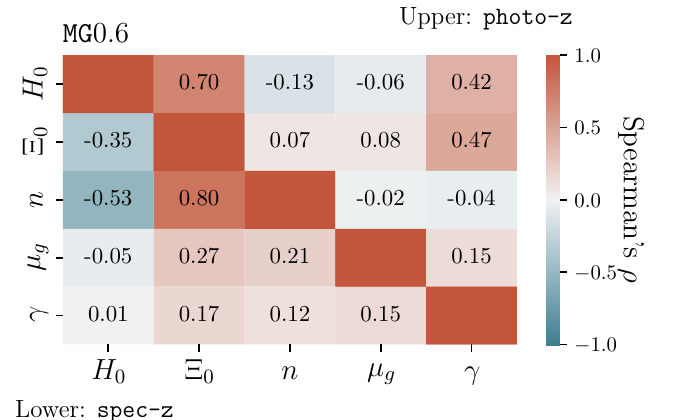}
\end{minipage}\hfill
\begin{minipage}{0.33\textwidth}
    \centering\includegraphics[width=\textwidth]{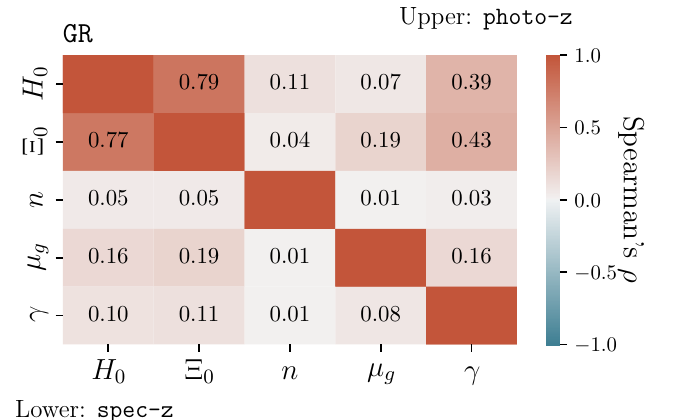}
\end{minipage}\hfill
\begin{minipage}{0.33\textwidth}
   \centering\includegraphics[width=\textwidth]{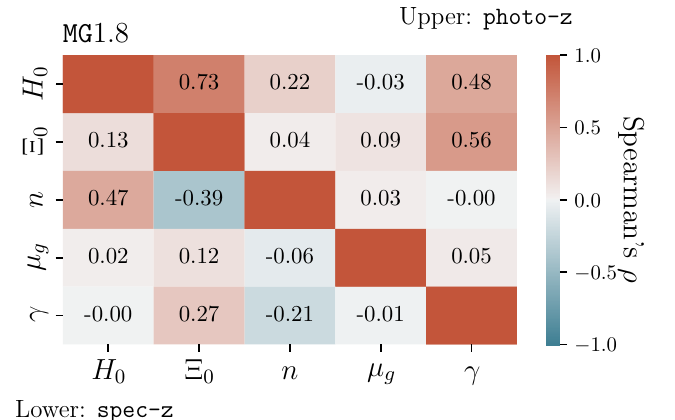}
\end{minipage}
    \caption{Correlation among selected hyperparameters in first MCMC configuration. The strength and direction of the correlation were quantified with a Spearman coefficient. In each panel, the upper corner plot shows the results in the \texttt{photo-z} case, while the lower triangle shows the results in the \texttt{spec-z} case. The left panel shows results from the \texttt{MG}0.6 GW catalog, the center panel from the \texttt{GR} catalog, and the right panel from the \texttt{MG}1.8 GW catalog.}
    \label{fig:par-correlation}
\end{figure*}

Similar results were obtained for the merger rate parameter $\gamma$, though the fiducial value lies slightly outside the 68\% C.L. for \texttt{MG}0.6 and \texttt{GR}.
Several factors may contribute to this: the \texttt{MG}1.8 catalog contains a higher density of events at low redshifts, potentially enhancing its constraining power; alternatively, the problem could be due to an insufficient parameter-space sampling in the current injection set. 
Nevertheless, since both $H_0$ and $\Xi_0$ are recovered without bias and exhibit a much weaker correlation with $\gamma$ than with each other (see next section and \Cref{fig:par-correlation}), this deviation does not affect our conclusions.
A more detailed investigation of this issue is left for future work.

\subsection{Hyperparameter correlations}
In \Cref{fig:table-2d-contours}, we show the correlations between cosmological, MG, and selected population hyperparameters across different catalogs and redshift error assumptions.
For the population hyperparameters, we focused on the mass peak position, $\mu_g$, and on the first slope of the merger rate, $\gamma$, as these are the most correlated with $H_0$ and $\Xi_0$.
To further quantify these correlations, we calculated the Spearman rank correlation coefficient ($\rho$) \citep{zwillinger1999crc}.
This metric, which ranges from $-1$ to $1$, quantifies the strength and direction of the correlation: a positive value indicates correlated variables, a negative value indicates a monotonically decreasing relation, and zero represents two non-correlated variables.
\Cref{fig:par-correlation} shows Spearman coefficients obtained when inferring all hyperparameters for the three GW catalogs.
The lower (upper) triangle of each panel displays the results obtained using spectroscopic (photometric) redshifts.

In the photometric case, all catalogs exhibit a strong correlation between $\Xi_0$ and $H_0$.
When spectroscopic redshifts are used, the degeneracy between these two hyperparameters is significantly reduced - except in the \texttt{GR} case. 
Furthermore, the constraints become much tighter due to the additional information provided by the spectroscopic galaxy catalog.

In the \texttt{MG}0.6 \texttt{spec-z}, $n$ is strongly correlated with $\Xi_0$ and anticorrelated with $H_0$.
These correlations reverse sign in the \texttt{MG}1.8 \texttt{spec-z} case.
In contrast, for the \texttt{GR} \texttt{spec-z} scenario, $n$ shows no significant correlation with either $H_0$ or $\Xi_0$.
We also note that in all \texttt{photo-z} cases, the Spearman coefficients of $n$ with $H_0$ and $\Xi_0$ are smaller than the in the respective \texttt{spec-z} cases; however this is due to the much weaker constraints on $n$.

When using photometric redshifts, we find that $\Xi_0$ is slightly correlated with the mass peak $\mu_g$ - particularly in the \texttt{GR} case - and strongly correlated with $\gamma$.
However, the $\Xi_0$-$\gamma$ correlation is notably weaker than when the galaxy catalog is not included (see, e.g., the corner plots 5 and 6 in \citealt{Mancarella:2021ecn}).
Remarkably, we find that using spectroscopic redshifts breaks these correlations.

\section{Conclusions}

 In this work, we presented an enhanced version of \CHIMERA, a code designed for Bayesian inference of cosmological, modified-gravity, and population hyperparameters using standard sirens and galaxy catalogs.
 This second release addresses the computational bottlenecks of the first version by including three distinct KDE algorithms used to compute the likelihood.
 We introduced these algorithms, detailing their differences in terms of results and computational efficiency, and validated their results against one another.
 As a next step, we plan to extend this validation test to real data and/or larger datasets to assess possible systematic effects. 
This will be one of the main goals of a future blinded-mock-data challenge between \CHIMERAsecond and similar pipelines.

Using \CHIMERAsecond, we forecasted constraints for an O5-like detector network in an optimistic scenario.
Three BBH populations have been studied: one assuming no modified GW propagation, one where the effective luminosity distance is greater than the electromagnetic one ($\Xi_0 = 1.8$) and one where it is smaller ($\Xi_0 = 0.6$).
Population hyperparameters are jointly inferred with MG and cosmological ones.
We studied both cases in which the galaxies have spectroscopic and photometric redshifts.
The fiducial MG and cosmological hyperparameters are recovered at 68\% C.L. in all scenarios. 
With spectroscopic redshift, the fiducial values of $\Xi_0$ can be distinguished across all scenarios, regardless of whether $H_0$ is marginalized or fixed.
However, the $H_0$ precision, which is always below $1\%$ when the MG hyperparameters are fixed, degrades by factors of $\sim2.1$, $\sim3.9$, and $\sim7.2$ when the MG hyperparameters are marginalized. 
Notably, the marginalization over MG hyperparameters strongly weakens the dependence of the $H_0$ precision on the number of golden sirens.
In the photometric case, the three $\Xi_0$ values can only be distinguished when $H_0$ is not inferred, and become indistinguishable when $H_0$ is marginalized.
In the photometric case, the constraints on $H_0$ degrade on average by a factor of $3.4$ ($1.2$) when the MG hyperparameters are marginalized (fixed) compared to the corresponding spectroscopic case.
In addition to this, using spectroscopic redshifts significantly reduces the correlation among cosmological, MG, and population hyperparameters.
This again highlights the importance of future spectroscopic surveys, such as WST \citep{WST:2024rai}, in terms of fully exploiting GWs as standard sirens for probing cosmology and modified GW propagation.
Lastly, the posterior distributions for $n$ remain nearly flat, and fixing $H_0$ does not improve the constraints on this parameter, both in the spectroscopic and photometric case.

We emphasize that these results are based on several simplifying assumptions.
First, we assume a complete galaxy catalog that includes only the most massive galaxies. In real galaxy surveys, however, galaxy-selection functions are more complex and must be properly accounted for in the cosmological analysis. At the same time, the completeness correction must be properly modeled, reflecting the probability of a galaxy hosting a merger event based on its astrophysical properties, rather than relying on uninformative assumptions such as uniform comoving volume completion. A more realistic and physically motivated framework that addresses these complexities will be developed in a future study.
While our current implementation uses this simplified framework, the results presented in this paper highlight the need for spectroscopic galaxy surveys and multiple GW detectors able to precisely localize GW sources in standard siren cosmology and GR testing.
Lastly, we plan to expand these forecasts to next-generation interferometers and prove how constraints will improve.
Although \CHIMERAsecond can handle approximately 50 times more events than \CHIMERAfirst, handling about $10^5$ events would require additional computational power. 
This could be achieved by parallelizing the likelihood calculation across multiple GPUs using \texttt{JAX} functionalities or the Message Passage Interface, which is a topic for future work.

\begin{acknowledgements}
    We thank Michele Mancarella for useful discussions and comments.
    We acknowledge the ICSC for awarding this project access to the EuroHPC supercomputer LEONARDO, hosted by CINECA (Italy).
    This material is based upon work supported by NSF's LIGO Laboratory which is a major facility fully funded by the National Science Foundation.
    MT acknowledges the funding from the European Union - NextGenerationEU, in the framework of the HPC project – “National Center for HPC, Big Data and Quantum Computing” (PNRR - M4C2 - I1.4 - CN00000013 – CUP J33C22001170001). MM acknowledges the financial contribution from the grant PRIN-MUR 2022 2022NY2ZRS 001 “Optimizing the extraction of cosmological information from Large Scale Structure analysis in view of the next large spectroscopic surveys” supported by NextGenerationEU. MM and NB acknowledge the financial contribution from the grant ASI n. 2024-10-HH.0 “Attività scientifiche per la missione Euclid – fase E”.

    We acknowledge the use of the following software: \texttt{NumPy} \citep{Harris:2020xlr}, \texttt{JAX} \citep{jax2018github}, \texttt{equinox} \citep{kidger2021equinox}, \texttt{matplotlib} \citep{Hunter:2007ouj}, \texttt{seaborn} \citep{Waskom:2021psk}, \texttt{arviz} \citep{arviz_2019}, \texttt{emcee} \citep{Foreman-Mackey:2012any},  \texttt{ChainConsumer} \citep{Hinton2016}, \texttt{GWFAST} \citep{Iacovelli:2022mbg}, \CHIMERAfirst \citep{Borghi:2023opd}.
\end{acknowledgements}

\bibliographystyle{aa}
\bibliography{references}

\begin{appendix} 

\section{Kernels comparison and validation}\label{app:kernel-validation}

In \Cref{fig:corner-validation} we compare the MCMC results obtained using three different kernels: the one that implements \CHIMERAfirst, the "many-1D" \eqref{eq:gw-kernel-many1d}, and the "single-1D" \eqref{eq:gw-kernel-single1d}.
The results are obtained using the ``O5-like'' catalog of \cite{Borghi:2023opd} and spectroscopic galaxy redshifts.
Overall, the posteriors obtained in the various cases are in excellent agreement.
\begin{minipage}{\textwidth}
    \centering
    \includegraphics[width=\textwidth]{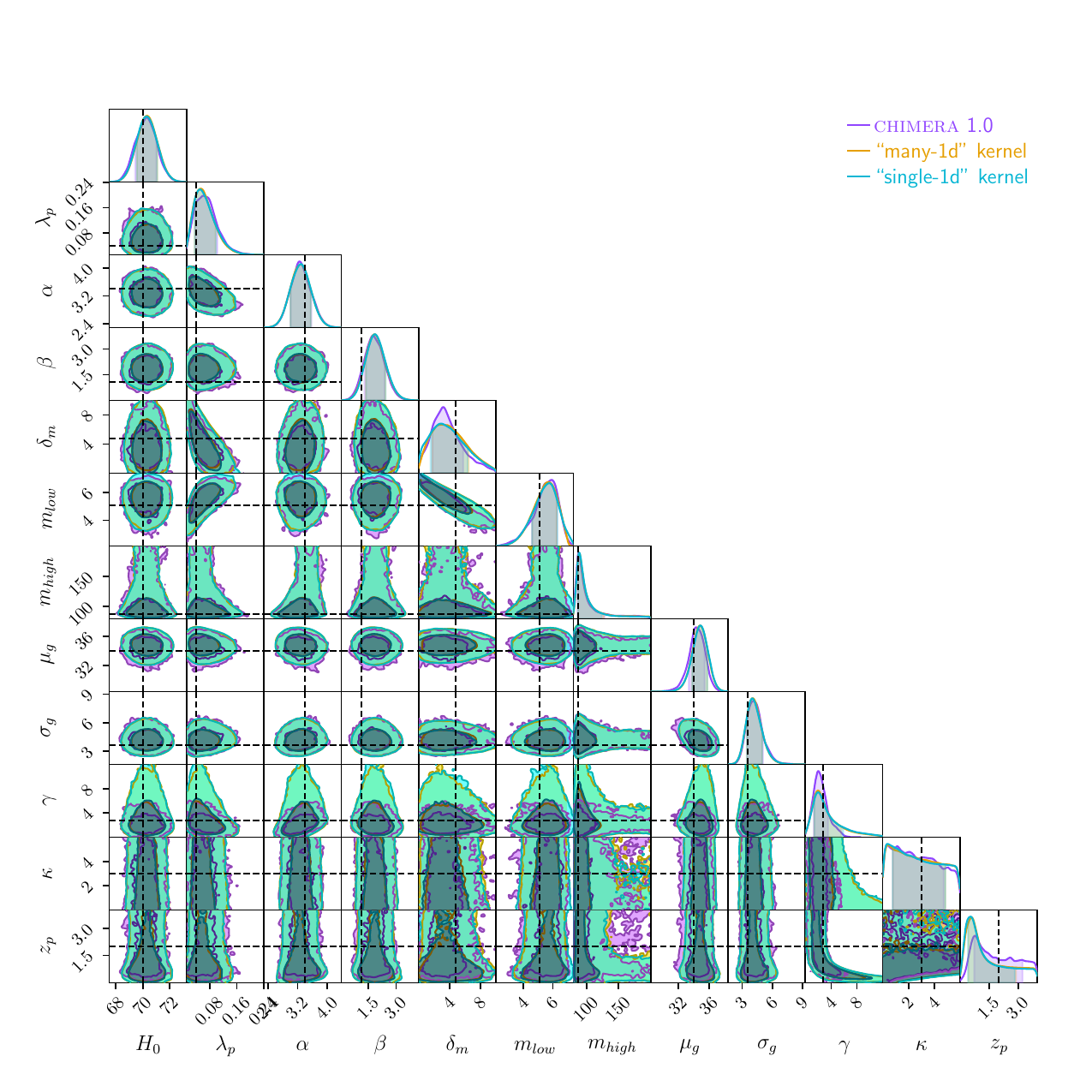}
    \captionof{figure}{Comparison between the MCMC results obtained using the kernels implemented in \CHIMERAsecond and the one implemented in \CHIMERAfirst. The contours represent the 68\% and 95\% C.L. The dotted lines indicate the hyperparameter fiducial values.}
    \label{fig:corner-validation}
\end{minipage}

\end{appendix}

\end{document}